\documentclass[referee]{svmult}

\usepackage[squaren, Gray, cdot]{SIunits}
\usepackage{mathptmx}       
\usepackage{helvet}         
\usepackage{courier}        
\usepackage{type1cm}        
\usepackage{color}
\usepackage{enumitem}
\usepackage{lscape}
\usepackage{wasysym}
\usepackage{supertabular}
\usepackage{multirow}
\usepackage{eurosym}
\usepackage{longtable}

\usepackage{makeidx}         
\usepackage{graphicx}        
\usepackage{multicol}        
\usepackage[bottom]{footmisc}

\makeindex             

\begin{document}

\title*{The Hera Saturn Entry Probe Mission}

\titlerunning{The Hera Saturn Entry Probe Mission}

\author{O. Mousis, D.H. Atkinson, T. Spilker, E. Venkatapathy, J. Poncy, R. Frampton, A. Coustenis, K. Reh, J.-P. Lebreton, L. N. Fletcher, R. Hueso, M. J. Amato, A. Colaprete, F. Ferri, D. Stam, P. Wurz, S. Atreya, S. Aslam, D. J. Banfield, S. Calcutt, G. Fischer, A. Holland, C. Keller, E. Kessler, M. Leese, P. Levacher, A. Morse, O. Mu\~noz, J.-B. Renard, S. Sheridan, F.-X. Schmider, F. Snik, J. H. Waite, M. Bird, T. Cavali\'e, M. Deleuil, J. Fortney, D. Gautier, T. Guillot, J. I. Lunine, B. Marty, C. Nixon, G. S. Orton, A. S\'anchez-Lavega}

\institute{O. Mousis, M. Deleuil, P. Levacher \at Aix Marseille Universit{\'e}, CNRS, LAM (Laboratoire d'Astrophysique de Marseille) UMR 7326, 13388, Marseille, France, \email{olivier.mousis@lam.fr}
\and D. H. Atkinson \at Department of Electrical and Computer Engineering, University of Idaho, Moscow, Idaho, USA
\and T. Spilker \at Solar System Science \& Exploration, Monrovia, USA
\and E. Venkatapathy, A. Colaprete \at NASA Ames Research Center, Moffett Field, California, USA
\and J. Poncy \at Thales Alenia Space, Cannes, France
\and R. Frampton \at The Boeing Company, Huntington Beach, California, USA
\and A. Coustenis, D. Gautier \at LESIA, Observatoire de Paris, CNRS, UPMC, Univ. Paris-Diderot, France
\and K. Reh, G. S. Orton \at Jet Propulsion Laboratory, California Institute of Technology, 4800 Oak Grove Dr., Pasadena, CA 91109, USA
\and J.-P. Lebreton \at LPC2E, CNRS-Universit{\'e} d'Orl{\'e}ans, 3a Avenue de la Recherche Scientifique, 45071 Orl{\'e}ans Cedex 2, France\\
LESIA, Observatoire de Paris, CNRS, UPMC, Univ. Paris-Diderot, France
\and L. N. Fletcher, S. B. Calcutt \at 
Atmospheric, Oceanic and Planetary Physics, Clarendon Laboratory, University of Oxford, Parks Road, Oxford OX1 3PU, UK
\and R. Hueso, A. S\'anchez-Lavega \at Departamento F\' isica Aplicada I, Universidad del Pa\'is Vasco UPV/EHU, ETS Ingenier\'ia, Alameda Urquijo s/n, 48013 Bilbao, Spain\\
Unidad Asociada Grupo Ciencias Planetarias UPV/EHU-IAA(CSIC), 48013 Bilbao, Spain
\and M. J. Amato, S. Aslam, C. A. Nixon \at NASA Goddard Space Flight Center, Greenbelt, MD 20771, USA
\and F. Ferri \at Universit\`a degli Studi di Padova, Centro di Ateneo di Studi e Attivit\`a Spaziali ``Giuseppe Colombo'' (CISAS), via Venezia 15, 35131 Padova, Italy
\and D. Stam \at Aerospace Engineering, Technical University, Delft, the Netherlands
\and P. Wurz \at Space Science \& Planetology, Physics Institute, University of Bern, Sidlerstrasse 5, 3012 Bern, Switzerland
\and S. Atreya \at Department of Atmospheric, Oceanic, and Space Sciences, University of Michigan, Ann Arbor, MI 48109-2143, USA
\and D. J. Banfield, J. I. Lunine \at Center for Radiophysics and Space Research, Space Sciences Building Cornell University,  Ithaca, NY 14853, USA
\and G. Fischer \at Space Research Institute, Austrian Academy of Sciences, Schmiedlstrasse 6, A-8042 Graz, Austria
\and A. Holland, A. Morse, S. Sheridan, M. Leese \at Department of Physical Sciences, The Open University, Walton Hall, Milton Keynes MK7 6AA, UK
\and C. Keller, F. Snik \at Leiden Observatory, Leiden University, P.O. Box 9513, NL-2300 RA Leiden, The Netherlands
\and E. Kessler \at Institute of Photonic Technology, Albert-Einstein-Str. 9, 07745 Jena, Germany
\and O. Mu\~noz \at Instituto de Astrof\'isica de Andaluc\'ia, CSIC, Glorieta de la Astronomia s/n, Granada 18008, Spain
\and J.-B. Renard \at LPC2E, CNRS-Universit{\'e} d'Orl{\'e}ans, 3a Avenue de la Recherche Scientifique, 45071 Orl{\'e}ans Cedex 2, France
\and F.-X. Schmider, T. Guillot \at Observatoire de la C\^ote d'Azur, Laboratoire Lagrange, BP 4229, 06304 Nice cedex 4, France
\and J. H. Waite \at Southwest Research Institute, San Antonio, TX 78228, USA
\and M. Bird \at University of Bonn, Bonn, Germany
\and T. Cavali\'e \at Max-Planck-Institut f\"ur Sonnensystemforschung, Justus von Liebig Weg 3, 37077 G\"ottingen, Germany 
\and J. Fortney \at University of California/Santa Cruz, California, USA
\and B. Marty \at CRPG-CNRS, Nancy-Universit\'e, 15 rue Notre Dame des Pauvres, 54501 Vandoeuvre-ls-Nancy, France
}

%
%
\maketitle

\abstract{The {\it Hera} Saturn entry probe mission is proposed as an M--class mission led by ESA with a contribution from NASA. It consists of one atmospheric probe to be sent into the atmosphere of Saturn, and a Carrier--Relay spacecraft. In this concept, the {\it Hera} probe is composed of ESA and NASA elements, and the Carrier--Relay Spacecraft is delivered by ESA. The probe is powered by batteries, and the Carrier--Relay Spacecraft is powered by solar panels and batteries. We anticipate two major subsystems to be supplied by the United States, either by direct procurement by ESA or by contribution from NASA: the solar electric power system (including solar arrays and the power management and distribution system), and the probe entry system (including the thermal protection shield and aeroshell). {\it Hera} is designed to perform {\it in situ} measurements of the chemical and isotopic compositions as well as the dynamics of Saturn's atmosphere using a single probe, with the goal of improving our understanding of the origin, formation, and evolution of Saturn, the giant planets and their satellite systems, with extrapolation to extrasolar planets. {\it Hera}'s aim is to probe well into the cloud-forming region of the troposphere, below the region accessible to remote sensing, to the locations where certain cosmogenically abundant species are expected to be well mixed. By leading to an improved understanding of the processes by which giant planets formed, including the composition and properties of the local solar nebula at the time and location of giant planet formation, {\it Hera} will extend the legacy of the Galileo and Cassini missions by further addressing the creation, formation, and chemical, dynamical, and thermal evolution of the giant planets, the entire solar system including Earth and the other terrestrial planets, and formation of other planetary systems.
}


\section{Introduction}
\label{sec:1}

The {\it Hera} Saturn entry probe mission consists of one atmospheric probe to be sent into the atmosphere of Saturn, and a Carrier-Relay Spacecraft (CRSC). {\it Hera} will perform {\it in situ} measurements of the chemical and isotopic compositions as well as the dynamics of Saturn's atmosphere using a single probe, with the goal of improving our understanding of the origin, formation, and evolution of Saturn, the giant planets and the solar system. {\it Hera} will probe well into the cloud-forming region of the troposphere, below the region accessible to remote sensing, to the locations where certain cosmogenically abundant species are expected to be well mixed.

The formation and evolution of the giant planets hold many keys to understanding the formation and evolution of the solar system as a whole, including the terrestrial planets, as well as exoplanetary systems. Key measurements include the composition and processes within giant planet atmospheres, gravitational fields, magnetospheres, and systems of moons. The {\it Galileo} probe provided {\it in situ} measurements of the chemical and isotopic composition of Jupiter's atmosphere. Of particular importance, the Jovian helium abundance was determined with a high accuracy. Moreover, the {\it Galileo} probe revealed unexpected enrichments of the noble gases Ar, Kr and Xe with respect to the solar abundances. Additionally, the {\it Galileo} probe mass spectrometer measured the $^{14}$N/$^{15}$N ratio, which strongly suggested that the nitrogen in Jupiter's atmosphere was acquired from the protosolar nebula (PSN). The {\it Galileo} probe and orbiter mission to Jupiter, complemented by the {\it Juno} mission currently en route to Jupiter and the L--class {\it JUICE} mission selected by ESA, will provide a solid understanding of the Jupiter system. The {\it Cassini} orbiter is providing valuable observations of Saturn's upper atmosphere, system of moons, gravitational field, and magnetosphere. However, the {\it Huygens} probe was destined to enter Titan's atmosphere and did not explore Saturn's atmosphere. 

The key missing element towards a similar system understanding of Saturn and an improved context for understanding the {\it Galileo}, {\it Juno}, and {\it JUICE} studies of Jupiter are the measurements of the composition and of the processes within Saturn's deeper atmosphere that only {\it in situ} exploration can provide. The {\it Hera} probe will use mass spectrometry to measure the abundances of hydrogen, helium, neon, argon, krypton, xenon, carbon, nitrogen, sulfur, and their compounds at near-equatorial latitude down to at least 10 bars. During its descent, {\it Hera} will also sample key isotopic ratios D/H, $^3$He$^/$4He, $^{20}$Ne/$^{21}$Ne/$^{22}$Ne, $^{36}$Ar/$^{38}$Ar, $^{12}$C/$^{13}$C, $^{14}$N/$^{15}$N, $^{16}$O/$^{17}$O/$^{18}$O, $^{82}$Kr/$^{83}$Kr/$^{84}$Kr/$^{86}$Kr, and $^{129}$Xe/$^{130}$Xe/$^{132}$Xe/$^{134}$Xe/$^{136}$Xe. {\it In situ} measurements of Saturn's well-mixed atmosphere gases will provide a vital comparison to the {\it Galileo} probe measurements at Jupiter, and a crucial ``ground truth'' for the remote sensing investigations by the {\it Cassini} orbiter. {\it Hera} will investigate Saturn's atmospheric dynamics along its descent trajectory, from (1) the vertical distribution of the pressure, temperature, clouds and wind speeds, and (2) deep wind speeds, differential rotation and convection, by combining {\it in situ} probe measurements and gravity and radiometric measurements from the carrier. {\it Hera} is the next logical step in our exploration of the Gas Giants beyond the {\it Voyager}, {\it Galileo} and {\it Cassini} missions.

{\it Hera} will lead to an improved understanding of the processes responsible for the formation of giant planets (contribution of the local solar nebula, accretion of icy planetesimals, and nature and formation temperature of the latter). The {\it Hera} data will shed light on the composition of giant planet precursors and on the dynamical evolution of the early solar system. {\it Hera} will also address the question as to why Jupiter and Saturn are so different in size, density and core dimension, investigating different pathways to planetary formation, thereby providing new insights on the mechanisms that led to the stunning diversity of giant planets.

{The {\it Hera} probe concept as proposed in response to ESA's Cosmic Vision Medium class size call in 2014 will be composed of ESA and NASA elements, and the CRSC will be delivered by ESA.} The probe will be powered by batteries. The CRSC will be powered by solar panels and batteries. We anticipated two subsystems to be supplied by the United States, either by direct procurement by ESA or by contribution from NASA: the solar electric power system (including solar arrays and the power management and distribution system), and the probe entry system (including the thermal protection shield and aeroshell). Following the highly successful example of the {\it Cassini}-{\it Huygens} mission, {\it Hera} would carry instruments from international partners, with scientists and engineers from both agencies and many affiliates participating in all aspects of mission development and implementation. A Saturn probe is currently one of the five missions on the NASA New Frontier's list, affirming that {\it Hera} science is a high priority for the European and American Planetary Science communities.

{\it Hera} flight could be with a Soyuz-Fregat launch from Kourou on a transfer trajectory to Saturn via several inner solar system flybys, with an arrival at Saturn 7-8 years after launch. The {\it Hera} CRSC releases the probe on a ballistic trajectory that will take it into Saturn's atmosphere a few weeks after its release. Prior to probe release, the CRSC would image Saturn to provide a global context for the probe science, as well as providing a local context of the probe entry location. Following the release of the {\it Hera} probe, the CRSC will be deflected to prepare for flight over the probe entry location for the probe data relay. 

The science objectives and measurement requirements of such a mission are described in Sec. \ref{sec:2}. The proposed science instruments are detailed in Sec. \ref{sec:3}. Section \ref{sec:4} is dedicated to a description of the current mission configuration and profile. We discuss the management scheme in Sec. \ref{sec:5}. Sec. \ref{sec:7} is devoted to summary and conclusion.

\section{Science Objectives And Requirements}
\label{sec:2}

\subsection{Context}
\label{sec:2.1}

The giant planets Jupiter, Saturn, Uranus and Neptune contain most of the mass and angular momentum of the sun's planetary system, and have played a significant role in shaping the solar system's large-scale architecture and evolution, as well as the properties of the smaller, inner worlds \cite{G05}. In particular, the formation of these planets has affected the timing and efficiency of volatile delivery to the Earth and other terrestrial planets \cite{CW01}. Understanding giant planet formation is therefore essential for understanding the origin and evolution of the Earth and other potentially habitable bodies within the solar system. The origin of the giant planets, their influence on the architecture of planetary systems, and the plethora of physical and chemical processes within their atmospheres, make the giant planets particularly important destinations for exploration. 

Both Jupiter and Saturn, the gas giants, are thought to have relatively small cores surrounded by massive envelopes composed primarily of hydrogen and helium. Uranus and Neptune are called ice giants because their density is consistent with the presence of a significant fraction of ices/rocks in their interiors. Despite the apparent grouping into two classes in the solar system, giant planets likely exist on a continuum, each carrying the characteristics of their particular formation environment. Comparative planetology of the sun's four giants is therefore essential to reveal the formational, migrational, and evolutionary processes during the early ages of the PSN.

The scientific goals of {\it Hera} are fully detailed in \cite{Mousis14}. The {\it in situ} exploration of Saturn's atmosphere addresses two themes that reach far beyond the unique knowledge gained about an individual planet, including (i) the formation history of the solar system and extrasolar planetary systems, and (ii) the processes that affect the vertical structure of temperatures, clouds and gaseous composition in planetary atmospheres. Examples of the latter are the stochastic and positional variances within the PSN, the depth of atmospheric zonal winds, the propagation of atmospheric waves, the formation of clouds and hazes, and disequilibrium processes of photochemistry and vertical mixing that are common to all planetary atmospheres, from terrestrial planets to gas and ice giants and from brown dwarfs to exoplanets.

\subsection{Why In Situ Measurements Are Needed}
\label{sec:2.2}

We have obtained most of our knowledge on the physical properties of the sun's giant planets through remote sensing from orbiters, fly-by missions, and ground-based telescopes. At visible wavelengths, {\it remote sensing} captures scattered and reflected sunlight, with a penetration depth into an atmosphere down to the upper hazes and clouds, At longer wavelengths, the thermal radiation from deeper layers emerges from the top of the planetary atmosphere. Indeed, important physical data addressing planetary composition, structure, and dynamics can be obtained with an orbiting spacecraft, as illustrated by the successful {\it Galileo} and {\it Cassini} missions. The information content of remote sensing data, however, remains severely limited due to (i) the degeneracies between the effects of temperatures, clouds, hazes, and gas abundances on the emergent spectra, and (ii) the limited penetration depth and vertical resolution.

As an example of the latter, the vertical distribution of many gases is strongly determined by chemical and condensation processes: many of the most common elements are locked away in a condensed phase, such as clouds or hazes in the deeper troposphere, hiding the main volatile reservoir from the reaches of remote sensing. The abundances of these gases in the upper atmospheric regions as derived from remote sensing data will thus not be representative of the bulk reservoir. Examples are NH$_3$ and H$_2$S (that will form NH$_4$SH clouds), H$_2$O, and other minor species such as PH$_3$, AsH$_3$, GeH$_4$ and tropospheric CO. Only by penetrating the ``visible'' weather layer, with the stratospheric hazes and upper clouds, can we sample the deeper troposphere and determine the true atmospheric composition. With {\it in situ} measurements, we will also be able to retrieve the vertical distribution of the lower tropospheric clouds and hazes, and the microphysical properties (size, shape, composition) of their particles that not only act as storage for elements, but also strongly influence the radiation field, the chemical and dynamical processes.

With {\it in situ} measurements, we can also trace the vertical dynamics that play a role in gas distributions. An example of the latter is the PH$_3$ profile, where the competing processes of photochemical sinks at high altitudes and sources from below could give a variety of profiles, depending on such factors as the strength of vertical upwelling. Also, a descending probe remains the only direct technique for measuring wind speeds at depths below the visible clouds. 

Some species such as the heavier noble gases do not leave distinct traces in spectra measured with remote sensing techniques and for these gases, {\it in situ} measurements are the only option to retrieve their abundances. A remarkable example of the capability of {\it in situ} probe measurements is illustrated by the exploration of Jupiter, where key data regarding the noble gases abundances and the helium mixing ratio could only be obtained through measurements by the {\it Galileo} probe \cite{Owen99}.

The {\it Galileo} probe measurements provided new insights into the formation of the solar system. In particular, the Jovian helium abundance was precisely determined with an accuracy of 2\% \cite{vZ98}, an accuracy impossible to achieve with remote sensing. An accurate measurement of the helium abundance in the atmospheres of giant planets is a key step towards understanding the fundamental problem of their formation and evolution in the solar and extrasolar systems. Moreover, the {\it Galileo} probe revealed the unexpected enrichments of Ar, Kr and Xe with respect to their solar abundances, suggesting different scenarios for Jupiter's formation. Another important result provided by the {\it Galileo} probe mass spectrometer was the $^{14}$N/$^{15}$N ratio, a value that suggested that Jupiter acquired its N$_2$ from the PSN reservoir.

The {\it Galileo} probe was designed to reach a depth of 10 bars, but survived to pressures exceeding 22 bars, descending into a region depleted in volatiles and gases by unusual ``hot spot'' meteorology \cite{Orton98,Wong04}. Therefore, the {\it Galileo} probe measurements of H$_2$O abundances are unlikely to represent Jupiter's bulk composition. The {\it Galileo} measurements nevertheless allowed us a giant step forward in understanding Jupiter. However, the solar-system chemical inventory and formation processes cannot be truly understood from the measured elemental and isotopic enrichments of a single giant planet. 

{\it In situ} exploration of the giant planets is the only way to completely characterize giant planet compositions in the solar system. In this context, a Saturn probe is the next natural step beyond {\it Galileo}'s {\it in situ} exploration of Jupiter \cite{Owen99}, the remote investigation of Jupiter's interior and gravity field by the {\it Juno} mission, and the {\it Cassini} spacecraft's orbital reconnaissance of Saturn.

\subsection{Measurement Priorities}
\label{sec:2.3}

An entry probe should reveal new insights into the vertical structures of temperatures, density, chemical composition and clouds during descent through a number of different atmospheric regions, from the stable upper/middle atmosphere to the convective troposphere. The probe would directly sample the condensation cloud decks and ubiquitous hazes whose composition, altitude and structure remain ambiguous due to the inherent limitations of remote sensing. In addition to bringing fundamental constraints on Saturn's formation conditions, {\it in situ} measurements would show how Saturn's atmosphere flows at a variety of different depths above, within and below the condensate clouds. The depth of probe penetration determines whether it can access the well-mixed regions for key condensable volatiles. In the present case, a shallow probe penetrating down to $\sim$10 bar would {\it in principle} sample NH$_3$ and H$_2$S both within and below their cloud bases, in the well-mixed regions of the atmosphere to determine the N/H and S/H ratios, in addition to noble gases and isotopic ratios. Note that the N determination could be a lower limit because ammonia is highly soluble in liquid water. Also, because the hypothesized water cloud is deeper than at least $\sim$12.6 bar in Saturn \cite{Atreya99}, the prospect of reaching the deep O/H ratio remains unlikely even if the probe survives beyond its design limit, unless a precise determination of the CO abundance (or any other species limited by reactions with the tropospheric water) is used to constrain H$_2$O/H$_2$. Nevertheless, measuring elemental abundances (in particular He, noble gases and other cosmogenically-common species) and isotopic ratios using a shallow entry probe on Saturn will provide a vital comparison to {\it Galileo}'s measurements of Jupiter, and a crucial ``ground-truth'' for the remote sensing investigations by the {\it Cassini} spacecraft. Table \ref{table1} ranks in order of priority the key {\it in situ} measurements that should be carried out by the {\it Hera} probe and its associated carrier.

\subsection{Required Instruments}
\label{sec:2.4}

The scientific requirements discussed above are addressed with a suite of scientific instruments located on the probe or the carrier spacecraft as detailed in Table \ref{table2}. At minimum, the science payload must contain two core instruments: a Mass Spectrometer (MS) and an Atmospheric Structure Instrument (ASI). These two instruments are sufficient to cover both Priority 1 and Priority 2 measurements. The MS will provide key measurements of the chemical and isotopic composition of Saturn's atmosphere, as well as its mean molecular weight. The key {\it in situ} measurements performed by the ASI will be the accelerometry during the probe entry phase and pressure, temperature and density profile during descent. A Radio Science Experiment (RSE), a Nephelometer, a Net Flux Radiometer (NFR) and a camera will address Priority 3 measurements. The RSE will include a Doppler Wind Experiment (DWE) dedicated to the measurement of the vertical profile of the zonal (east-west) winds along the probe descent path. It will also include an element dedicated to absorption measurements, to indirectly infer the abundance of ammonia. The Nephelometer will be devoted to the investigation of the composition and precise location of cloud layers. The NFR will measure the thermal profile and radiative energy balance in the atmosphere. A camera located on the carrier will provide i) contextual imaging of the probe entry site and ii) global characterization of Saturn's atmosphere at the time of probe entry. The Science Traceability Matrix is represented in Table \ref{table3}. 

\subsection{Probe Entry Zone}
\label{sec:2.5}
In the present paper, the trajectory selection is based on the selected carrier option, launch vehicle (Soyuz) capabilities, and the available probe thermal protection capability. The interplanetary trajectory and the probe entry location are inseparably linked.  Saturn's extensive ring system presents a severe collision hazard to an inbound probe. For various declinations of the spacecraft's approach asymptote, some latitudes are inaccessible because the trajectories to deliver to those latitudes would impact the rings.  Also, although it is possible to adjust the inclination of the approach orbit for purposes of accessing desired latitudes, this approach can greatly increase the atmosphere-relative entry speeds, possibly driving the mission to an expensive heat shield material technology development (see Sec. \ref{sec:4}). During the ESA assessment study, the issues of probe entry locations, approach and entry trajectories, and probe technologies will have to be treated together. 

With a single entry probe, the selected entry site must be carefully studied. {\it Saturn's equatorial zone is one potential site from the scientific point of view for a single entry probe} because of its meteorological activity that combines the emergence of large-scale storms \cite{SL91}, vertical wind shears in the troposphere \cite{GM11}, and upwelling enhanced volatiles and disequilibrium species \cite{F11,F09}. However, this may not be typical of Saturn's atmosphere, so detailed trades would need to be discussed during the study phase. Eastward jets (particularly the anticyclonic branch of eastward jets) located at equator might be good locations to retrieve the deep values of volatiles at higher levels in the atmosphere \cite{R09}. A primary requirement is that {\it volatile-depleted regions must be avoided for the entry site}. These zones are probably located at the cyclones in both poles and may also be located at the so-called ``storm-alley'' (mid-latitude regions of low static stability able to develop updrafts and downdrafts). In any case, there are several potential entry points and a decision where to enter must also be guided by the design of the thermal protection system of the probe. Envisaging {\it in situ} measurements in the equatorial region of Saturn appears to be the best compromise between science and engineering.

\section{Proposed Science Instruments}
\label{sec:3}

The {\it Hera} Saturn Probe mission will conduct {\it in situ} measurements of the structure, composition and fundamental processes operating within Saturn's atmosphere. Measurements will be made by a suite of instruments on the probe as it descends for up to 75--90 minutes under a parachute from the tropopause near 100 mbar, through the upper cloud decks, down to at least 10 bars. The Tier 1 instruments, designed to address the highest priority science goals, include a Mass Spectrometer and an Atmospheric Structure Instrument. The instruments comprising the Tier 2 payload address lower priority science goals, and include a Net Flux Radiometer, a Nephelometer, and a Radio Science experiment. While most instruments are located on the {\it Hera} probe itself, one ultrastable oscillator for the Radio Science experiment will be mounted on the Carrier. The Carrier will also have a camera operating at visible wavelengths to provide contextual imaging of the probe's entry site and a global characterization of Saturn's atmosphere at the entry time.

All instruments can operate on both the day and night side of Saturn, although the visible channel of the Net Flux Radiometer can only measure the altitude profile of solar energy absorption if the descent is on the dayside. The following section provides the investigation and measurement objectives of each instrument, including the measurement principle, the description of the instrument design, the resource requirements including mass, power, volume, and data rate, interface and calibration requirements, and a summary of technology readiness, heritage, and critical issues (if any). The total data returned from the probe will range from 30 to 40 megabits total. Table \ref{Inst} summarizes the main characteristics of the instruments (size, mass, power requirement, data rate and volume).

\subsection{Hera Mass Spectrometer}
\label{sec:3.1}

\subsubsection{Investigation Overview}
\label{sec:3.1.1}

The chemical, elemental, and isotopic composition of Saturn's atmosphere and its profile down to the 10 bar level will give important clues about the solar nebula at the location of Saturn's formation, about the formation of giant planets (in comparison to Jupiter) and Saturn's evolution to present state. Also, measurement of the composition of Saturn's atmosphere will provide details of the chemical structure of the atmosphere over the descent trajectory, which is not accessible to remote sensing investigations.

The measurement objective of the {\it Hera} Mass Spectrometer (MS) is to provide {\it in situ} measurements of the chemical, elemental, and isotopic composition of Saturn's atmosphere, and dependence of composition on pressure/altitude along the descent trajectory of the entry probe. The primary objective is the determination of the abundances of the major chemical species CH$_4$, NH$_3$, H$_2$O, H$_2$S, the He/H ratio, and the abundance of the noble gases Ne, Ar, Kr and Xe. Secondary objectives include isotopic ratios of major elements like H, He, C, and N, the abundances of minor chemical species, and the isotopic abundances of noble gases. Tertiary objective are the abundance of the oxygen isotopes. 

There is plenty of heritage for measuring the chemical composition during a descent through the atmosphere, for example the {\it Galileo} Probe mass spectrometer system \cite{Niemann96} or the {\it Huygens} Gas Chromatograph mass spectrometer \cite{Niemann02}. Over the last two decades, Time-Of-Flight Mass Spectrometers (TOF-MS) have been developed for space research, for example on the $Rosetta$ mission \cite{Balsiger07}, which have several advantages over the quadrupole mass spectrometers used before. For example, a TOF-MS is over 1000 times more sensitive than the {\it Cassini} INMS (ten times from ion source efficiency and 100 times from better duty cycle). Also for the gas inlet system and the gas enrichment system there is plenty of heritage from previous missions, again the {\it Galileo} Probe mass spectrometer system, the {\it Huygens} Gas Chromatograph mass spectrometer, or more recently the mass spectrometer experiments on the $Rosetta$ lander.

\subsubsection{Measurement Principle}
\label{sec:3.1.2}

The core of the {\it Hera} MS is a TOF-MS. Such instruments have several advantages for space research: i) all masses are measured at the same time, thus there is no need for scanning the mass, leading to an increase of the sensitivity, ii) they are simple and robust instruments very suitable for remote operation on a spacecraft, iii) they can be built light-weight. The cadence of mass spectrometric measurements is variable, from mass spectra accumulated every 1-second to integration up to 300 seconds. At suitable times, measurements of atmospheric gas are replaced by measurements of calibration gas, and measurements of gas enriched and separated from the bulk atmosphere. 

The atmospheric gas will enter the experiment via a gas inlet system comprising several independent entrances of various conductances, which will cover the pressure range of 0.1--10 bar level. The cadence of mass spectra is adjusted such that the vertical resolution is about 1.8 km along descent trajectory, which amounts to a total of about 400 mass spectra along the descent trajectory. 

Not all gases can be measured directly in the gas entering from the atmosphere, at least not with the desired accuracy. Noble gases, for example, will be separated from the entering gas and collected by a cryotrap enrichment system. After sufficient enrichment of the noble gases is accomplished they are released to the TOF-MS for a dedicated mass spectrometric measurement while the direct sampling of the atmosphere is interrupted. Similarly, the use of an additional cryotrap for the enrichment of hydrocarbons and other trace species will also be analyzed at regular intervals. 

The accuracy of some composition measurements will be enhanced by carrying several reservoirs of reference gases with an accurately known gas mixture. For example, for the measurement of the He/H ratio a gas container with a calibrated He/H mixture is part of the {\it Hera} MS experiment that will allow for the measurement of this ratio with an accuracy of 2\% or better. Similarly, a container with a calibrated mixture of noble gases, and a container with reference gases for key isotopes (H, C, N, and O), are included in the {\it Hera} Mass Spectrometer. 

\subsubsection{Design Description / Operating Principle}
\label{sec:3.1.3}

The {\it Hera} Mass Spectrometer consists of four units: the TOF-MS, the Tunable Laser Spectrometer (TLS), the Gas Separation and Enrichment System (GSES), and the Reference Gas System (RGS). 

The TOF-MS consists of a pulsed ion source, a time-of-flight drift path, an ion mirror (reflectron), and a fast ion detector. The TOF-MS is a compact instrument and has a mass range of 1--1000 u/e, a mass resolution of $m$/$\Delta$$m$ = 1100, and a very high sensitivity \cite{W12}. Ions, continuously generated in the ion source, are pulse-extracted, and sent as ion packets along the TOF path with a repetition frequency of 10 kHz, to the detector resulting in a mass spectrum. These spectra are accumulated for a defined integration period (1--300 seconds), depending on the desired vertical resolution along the descent trajectory. The integration of many spectra provides for a dynamic range of 6--7 decades in each accumulated spectrum, together with various detector gain steps and the gas enrichments at a dynamic range that exceeds 12 decades is achieved. 

The Tunable Laser Spectrometer (TLS) \cite{Durry02} will be employed as part of the {\it Hera} MS to measure the isotopic ratios with high accuracy of the molecules H$_2$O, NH$_3$, CH$_4$, CO$_2$ and others. TLS employs ultra-high spectral resolution tunable laser absorption spectroscopy ($\Delta\nu$ = 0.0005 cm$^{-1}$) in the near infra-red (IR) to mid-IR spectral region. TLS is a direct, non-invasive, and simple technique that for small mass and volume achieves sensitivities at the sub-ppb level for gas detection. Species abundances can be measured with accuracies of a few percent, and isotope determinations have an accuracy of about 0.1\%. With the TLS system one can derive the isotopic ratios of D/H, $^{13}$C/$^{12}$C, $^{15}$N/$^{14}$N, $^{18}$O/$^{16}$O, and $^{17}$O/$^{16}$O. A recent use of a TLS system was in the Sample Analysis at Mars (SAM) GC-MS system on Mars Science Laboratory (MSL) \cite{Webster11,Mahaffy12}. The GSES consists of a cryotrap, an ion pump, and a Non-Evaporable Getter (NEG), which together are used to achieve the noble gas enhancement. The NEG removes all constituents except methane and the noble gases. The cryotrap traps the products of the NEG process, except for helium and some neon. The ion pump then operates to pump away the helium, which is the second most abundant species in Saturn's atmosphere, thus enhancing the signal to noise ratio in the$^{17}$O/$^{16}$O remaining noble gases by about 200 times. This enrichment cell will be accessed periodically during descent to allow the noble gases to be analysed. The cryotrap for minor species will have a separate gas inlet. It will be heated periodically and a valve opened to allow the descent measurements to be interrupted for analysis.

The Reference Gas System consists of a central manifold and pressure sensor connected to the mass spectrometer via a capillary leak. Reference gas mixtures are stored in stainless steel 1 ml containers at a pressure of approximately 1 bar. Each reference gas will be admitted into the manifold by opening a single valve in a short pulse. These valves have a leak rate of less than 10$^{-8}$ mbar l s$^{-1}$ and a controllable pulse width of less than 1 ms; they are a development from Ptolemy $Rosetta$ heritage (TRL 5). Alternative valves are the same as used on Philae (TRL 9) but have a higher power requirement and a longer operating cycle of several minutes. 

The baseline proposal includes 3 reference gas mixtures; a hydrogen/helium mixture, a noble gas mixture and an isotope mixture. The composition of the isotopic reference gas will be a mixture of relatively inactive molecules, e.g. methane, carbon monoxide and nitrogen, depending upon the scientific targets. 

\begin{figure}
\begin{center}
\includegraphics[scale=1.5]{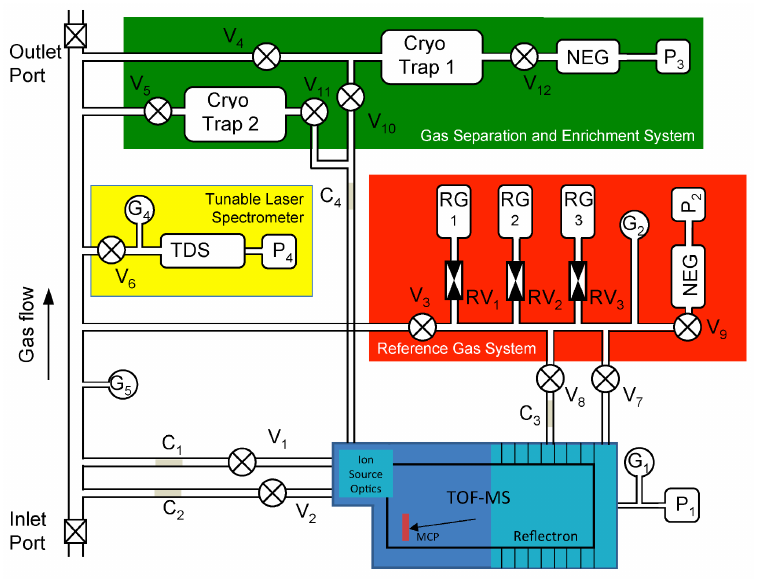}
\caption{Schematics of the Hera MS experiment, with the TOF-MS, the GSES, the RGS, and the TDS units. The elements are valves (V), regulating valves (RV), pressure gauges (G), conductance limiters (C), pumps (P), gas reservoirs (RG), and non-evaporative getter (NEG).}
\label{Hera_MS}      
\end{center}
\end{figure}

The RGS includes an ion pump and non-evaporable getter to remove gases between analyses and allow calibration of the mass spectrometer during cruise, a few hours before atmospheric entry and during the atmospheric descent. The ion pump adds a significant mass to the RGS, which could be reduced by using the GSES pump instead; however this adds to the complexity in the timing between the two systems and potentially results in cross contamination between the reference gases and the atmospheric samples (see Fig. \ref{Hera_MS}).

\subsection{Hera Atmospheric Structure Instrument}
\label{sec:3.2}

\subsubsection{Investigation overview}
\label{sec:3.2.1}

The {\it Hera} Atmospheric Structure Instrument ({\it Hera}-ASI) will make {\it in situ} measurements during entry and descent into the Saturn's atmosphere in order to investigate the atmospheric structure and dynamics, and electricity. The {\it Hera}-ASI scientific objectives are the determination of the atmospheric vertical pressure and temperature profiles, the evaluation of the density along the Probe trajectory and the investigation of the atmospheric electricity (e.g. lightning) by {\it in situ} measurements. {\it Hera}-ASI data will also contribute to the analysis of the atmospheric composition. Moreover, {\it Hera}-ASI will have a primary engineering function by establishing the entry trajectory and the probe altitude and vertical velocity profile for correlating all probe experiment data and to support the analysis of the Radio Science / Doppler Wind Experiment (DWE).

{\it In situ} measurements are essential for the investigation of the atmospheric structure and dynamics. {\it Hera} ASI will measure the atmospheric state ($P$, $T$ and density) as well as constraining atmospheric stability and dynamics, and the effect on atmospheric chemistry. The estimation of the temperature lapse rate can be used to identify the presence of condensation and eventually clouds, and to distinguish between saturated and unsaturated, stable and conditionally stable regions. The vertical variations in the density, pressure and temperature profiles provide information on the atmospheric stability and stratification, on the presence of winds, thermal tides, waves and turbulence in the atmosphere. 

{\it Hera} ASI will measure properties of Saturn lightning, determine the conductivity profile of the Saturnian troposphere, and detect the atmospheric Direct Current (DC) electric field. Atmospheric storm systems on Saturn with typical sizes of 2000-km \cite{Dyudina07} produce superbolt-like lightning discharges with energies up to 10$^{10}$ J \cite{Dyudina13}. To date the strong Saturn lightning radio emissions have only been measured from outside Saturn's ionosphere, i.e. mostly at frequencies $>$1 MHz and occasionally down to a few hundred kHz. Hence {\it Hera} ASI will measure the unknown lightning spectrum in the frequency range of $\sim$1--200 kHz, and obtain burst waveforms with different temporal resolutions and durations. A Saturn lightning flash typically lasts $\sim$100 ms and consists of many sub-discharges of the order of 0.1 ms, so waveforms over 100 ms with 0.1 ms resolution for the full flash and waveforms over 0.5 ms with 2$\mu$s resolution for the sub-strokes would be a sensible choice. The latter requires a sampling frequency of 500 kHz, which is also sufficient for obtaining the spectrum up to 200 kHz. Atmospheric conductivity and the DC electric field are important basic parameters of atmospheric electricity which provide indirect information about galactic cosmic ray ionization, aerosol charging inside and outside of clouds, properties of potential Schumann resonances and so on.

The proposed instrument will benefit from a strong heritage of the Huygens ASI experiment of the $Cassini$/$Huygens$ mission \cite{Fulchignoni02} and $Galileo$, and $Pioneer$ Venus ASI instruments \cite{Seiff92,Seiff80}.

\subsubsection{Measurement Principle}
\label{sec:3.2.2}

The key {\it in situ} measurements will be atmospheric density, pressure and temperature profile by measuring deceleration of the entry vehicle and performing direct temperature and pressure measurements during the descent phase \cite{Fulchignoni05,Seiff98}. Densities will be determined using measurements of the deceleration of the probe during entry. The flight profile of the probe, including variations in speed and angle of attack provide information regarding turbulence and vertical motions.  Once the probe heat shield is jettisoned, direct measurements of pressure, temperature and electrical properties will be performed. {\it Hera} ASI will monitor the acceleration experienced by the probe during the whole descent phase and will provide the unique direct measurements of pressure, temperature, conductivity, and DC electric field through sensors having access to the atmospheric flow.

\subsubsection{Design Description / Operating Principle}
\label{sec:3.2.3}

The {\it Hera} Atmospheric Structure Instrument (ASI) consists of several sensors both internal and external to the pressure vessel, and operates during high speed entry in the upper atmosphere and in descent when the probe is subsonic.  The proposed instrument design leverages strongly from the {\it Huygens} ASI experiment of the {\it Cassini}/{\it Huygens} mission \cite{Fulchignoni02} and the {\it Galileo} and Pioneer Venus ASI instruments \cite{Seiff92,Seiff80}. The {\it Hera} ASI consists of four primary sensor packages: (i) a three axial accelerometer (ASI-ACC), (ii) a Pressure Profile Instrument (ASI-PPI), (ii) temperature sensors (ASI-TEM) and (iv) an Atmospheric Electricity Package (ASI-AEP). The control, sampling and data management of the ASI sensors is handled by a central Data Processing Unit (DPU) including the main electronics for the power supply and conditioning, input/output and sensor control. The ASI-DPU interfaces directly to the entry probe processor.

The ASI-ACC will start to operate prior to the beginning of the entry phase, sensing the atmospheric drag experienced by the entry vehicle. Direct pressure and temperature measurement will be performed by the sensors having access to the atmospheric flow from the earliest portion of the descent until the end of the probe mission at approximately 10 bars. AEP will measure the atmospheric conductivity and DC Electric field in order to investigate the atmospheric electricity and detecting lighting.

\paragraph{\it Accelerometers}

The ACC package consisting of 3-axis accelerometers should be placed as close as possible to the center of mass of the entry vehicle. The main sensor is a highly sensitive servo accelerometer aligned along the vertical axis of the Probe, with a resolution of   10$^{-5}$ to 10$^{-4}$ m/s$^2$ (depending on the resolution setting) with an accuracy of 1\%. Accelerations can be measured in the 0-200 g range (where g is the Earth's acceleration of gravity). The Huygens servo accelerometer is the most sensitive accelerometer ever flown in a planetary entry probe \cite{Zarnecki04}. Having a triaxial accelerometer (namely one sensor located along each probe axis) will allow for an accurate reconstruction of the trajectory and attitude of the probe, and to sense the atmospheric drag in order to derive the entry atmospheric density profile. Assuming the HASI ACC Servo performance at Titan, a noise performance of some 0.3 $\mu$g is expected. The exact performance achievable, in terms of the accuracy of the derived atmospheric density, will also depend on the probe ballistic coefficients, entry speed and drag coefficient, all of which will differ somewhat from the Titan case.

\paragraph{\it Pressure Profile Instrument}

The ASI-PPI will measure the pressure during the entire descent with an accuracy of 1\% and a resolution of 1 micro bar. The atmospheric flow is conveyed through a Kiel probe inside the Probe where the transducers and related electronic are located. 

The transducers are silicon capacitive sensors with pressure dependent dielectricum. The pressure sensor contains a small vacuum chamber as dielectricum between the two electrode plates, where the external pressure defines the distance of these plates. Detectors with diaphragms of different pressure sensitivity will be utilized to cover the pressure range to $\sim$10 bar. The pressure is derived as a frequency measurement (within 3--20 kHz range) and the measurements is internally compensated for thermal and radiation influences.

\paragraph{\it Temperature Sensors}

The Temperature Sensors (TEM) utilize platinum resistance thermometers to measure the kinetic temperature during the descent just as in the {\it Huygens} Probe ASI and {\it Galileo} probe. Two thermometers are exposed to the atmospheric flow and effectively thermally isolated from the support structure. Each thermometer includes two redundant sensing elements: the primary sensor directly exposed to the airflow and a secondary sensor embedded into the supporting frame with the purpose to be used as spare unit in case of damage of the primary.  The principle of measurement is based on the variation of the resistance of the metallic wire with temperature. The reading of the thermometer is made by resistance comparison with a reference resistor, powered by a pulsed current.

TEM has been designed in order to have a good thermal coupling between the sensor and the atmosphere and to achieve high accuracy and resolution. Over the temperature range of 60--330 K these sensors maintain an accuracy of 0.1 K with a resolution of 0.02 K.

\paragraph{\it Atmospheric Electricity Package}

The Atmospheric Electricity Package (AEP) consists of sensors and a signal processing unit. Since Saturn's lightning is very intense and localized, it should be detectable by a short electric monopole, dipole, loop antenna or double probe from distances of several thousands of kilometers. The conductivity of the atmosphere can be measured with a mutual impedance probe. A current pulse is sent through the surrounding medium and the resulting voltage is measured by two passive electrodes from which the impedance of the medium can be determined. This can be corroborated by determining the discharge time (relaxation) of two charged electrodes. After the discharge, the natural DC electric field around the probe can also be measured with them. The signal processing unit (to be accomodated into the ASI main central unit) will amplify the signals, extract waveforms of bursts with different durations and temporal resolutions, perform spectral analysis at various frequency ranges (1--200 kHz or in the TLF below 3 Hz to detect Schumann resonances), and to provide active pulses and sensor potential control to handle the conductivity and DC electric field measurements.

\subsection{Hera Net Flux Radiometer Experiment}
\label{sec:3.3}

\subsubsection{Investigation Overview}
\label{sec:3.3.1}

\begin{figure}
\begin{center}
\includegraphics[scale=1.5]{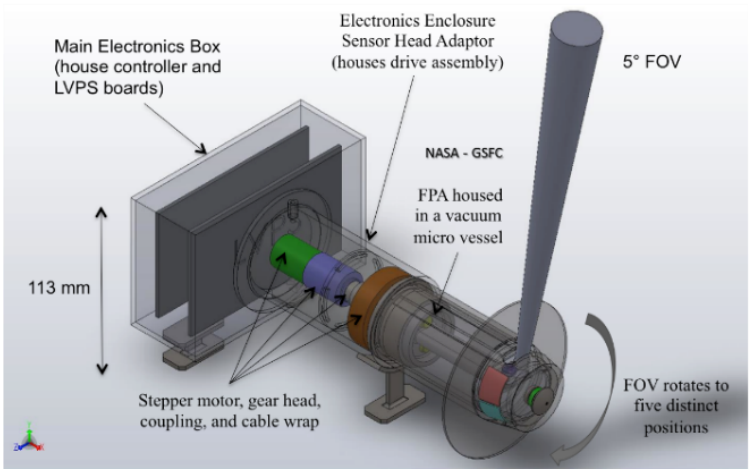}
\caption{NFR instrument concept showing a 5$\degree$ field-of-view that can be rotated by a stepper motor into five distinct look angles (NASA GSFC).}
\label{NFR1}      
\end{center}
\end{figure}

Two notable Net Flux Radiometer (NFR) instruments have flown in the past namely, the Large probe Infrared Radiometer (LIR) \cite{Boese80} on the Venus Probe, and the NFR on the {\it Galileo} Probe \cite{Sromovsky98} for {\it in situ} measurements within Venus and Jupiter's atmospheres, respectively. Both instruments were designed to measure the net radiation flux and upward radiation flux within their respective atmospheres as the Probe descended by parachute. The NASA GSFC Net Flux Radiometer (see Fig. \ref{NFR1}), builds on the lessons learned from the {\it Galileo} Probe NFR experiment and is designed to determine the net radiation flux within Saturn's atmosphere. The nominal measurement regime for the NFR extends from $\sim$0.1 bar to at least 10 bars, corresponding to an altitude range of $\sim$79 km above the 1 bar level to $\sim$154 km below it. These measurements will help to define sources and sinks of planetary radiation, regions of solar energy deposition, and provide constraints on atmospheric composition and cloud layers.
 
The primary objective of the NFR is to measure upward and downward radiative fluxes to determine the radiative heating (cooling) component of the atmospheric energy budget, determine total atmospheric opacity, identify the location of cloud layers and opacities, and identify key atmospheric absorbers such as methane, ammonia, and water vapor. The NFR measures upward and downward flux densities in two spectral channels. The specific objectives of each channel are: 

\begin{itemize}
\item Channel 1 (solar, 0.4-to-5$\mu$m). Net flux measurements will determine the solar energy deposition profile; upward flux measurements will yield information about cloud particle absorption and scattering;
\item Channel 2 (thermal, 4-to-50$\mu$m). Net flux measurements will define sources and sinks of planetary radiation. When used with calculations of gas opacity effects, these observations will define the thermal opacity of particles.
\end{itemize}

\subsubsection{Measurement Principle}
\label{sec:3.3.2}

The NFR measures upward and downward radiation flux in a 5$\degree$ field-of-view at five distinct look angles, i.e., $\pm$80$\degree$, $\pm$45$\degree$, and 0$\degree$, relative to zenith/nadir. The radiance is sampled at each angle approximately once every $\sim$2s.  

\begin{figure}
\begin{center}
\includegraphics[scale=1.4]{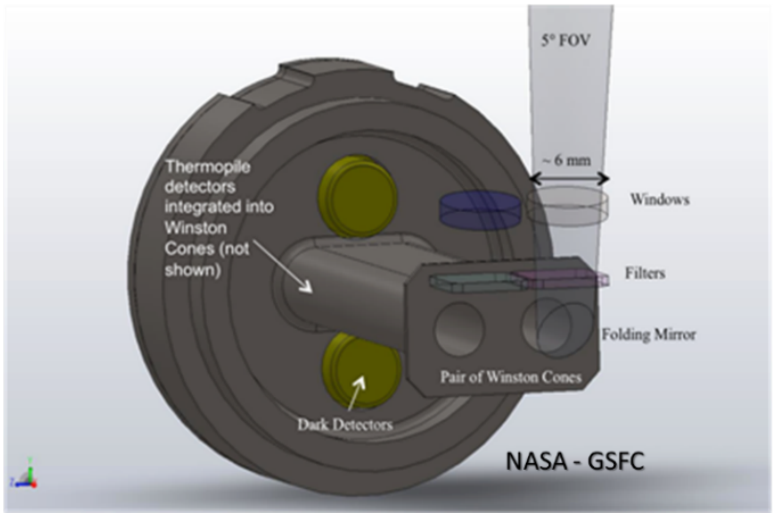}
\caption{NFR Focal Plane Assembly showing construction of Winston cones to limit FOV in each channel (NASA GSFC).}
\label{NFR2}      
\end{center}
\end{figure}

The NFR Focal Plane Assembly (FPA), Figure \ref{NFR2}, is comprised of bandpass filters, folding mirrors, non-imaging Winston cone concentrators, and radiation hard uncooled thermopile detectors housed in a windowed vacuum micro-vessel that is rotated to the look angle by a stepper motor. 

Assuming a thermopile voltage responsivity of 295 V/W, an optical efficiency of 50\%, a detector noise of 18 nV/$\sqrt{\rm Hz}$ and an Application Specific Integrated Circuit (ASIC) input referred noise of 50 nV/$\sqrt{\rm Hz}$ with 12-bit digitization gives a system signal-to-noise ratio of 300 to 470 in the solar spectral channel and 100 to 12800 in the thermal spectral channel for atmospheric temperature and pressure ranges encountered in the descent, i.e., 80 to 300 K and 0.1 to 10 bar respectively.

\subsubsection{Design Description / Operating Principle}
\label{sec:3.3.3}

\begin{figure}
\begin{center}
\includegraphics[scale=1.4]{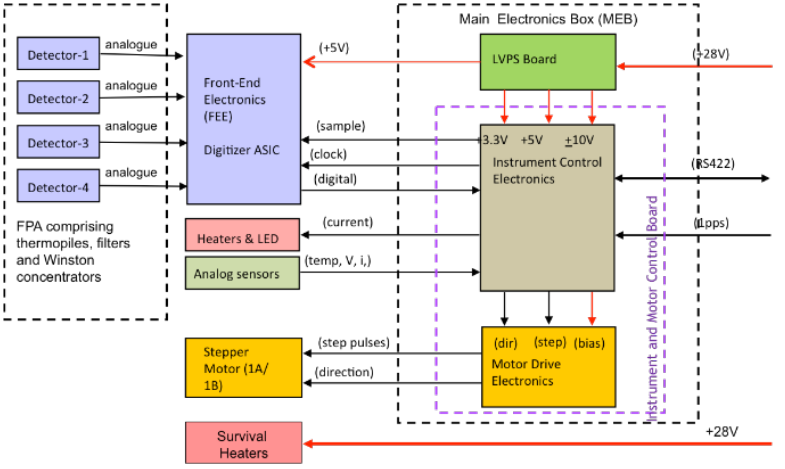}
\caption{Block diagram of the NFR showing the major subsystems and Probe interfaces. The redundant features are not shown (NASA GSFC).}
\label{NFR3}      
\end{center}
\end{figure}

\begin{figure}
\begin{center}
\includegraphics[scale=1.4]{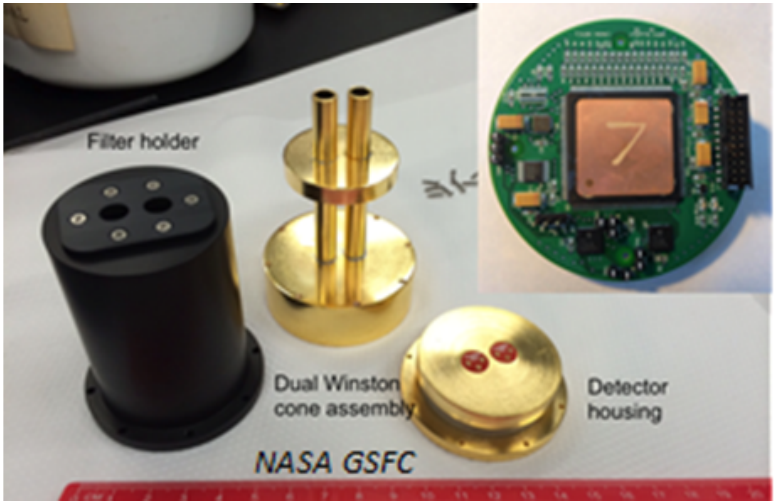}
\caption{NASA GSFC is testing early engineering models of the critical components the NFR. Dual Winston cone assembly and thermopile FEE readout (diameter $\sim$70mm) that uses a GSFC rad-hard mixed-signal ASIC (NASA GSFC).}
\label{NFR4}      
\end{center}
\end{figure}

A physical and functional block diagram of the NFR is shown in Fig. \ref{NFR3}. The focal plane consists of four single pixel thermopile detectors (solar, thermal and two dark channels), bandpass filters and Winston concentrators. The Front End Electronics (FEE) readout, see inset of Fig. \ref{NFR4}, uses a custom radiation-hardened-by-design mixed-signal ASIC for operation with immunity to 174 MeV-cm$^2$/mg single event latch-up and 50 Mrad (Si) total ionizing dose \cite{Quiligan14}. The ASIC has sixteen low-noise chopper stabilized amplifier channels that have configurable gain/filtering and two temperature sensor channels that multiplex into an on-chip 16-bit sigma-delta analog-digital converter (SDADC). The ASIC uses a single input clock ($\sim$1.0 MHz) to generate all on-chip control signals such as the chopper/decimation clocks and integrator time constants.  The ASIC also contains a radiation tolerant 16-bit 20 MHz Nyquist ADC for general-purpose instrumentation needs. 

The Main Electronic Box (MEB) is a redundant electrical system for science and housekeeping telemetry and thermal sensing and control. The two main elements of the MEB are the instrument and motor control board (comprising the instrument control and the motor drive electronics) and the Low Voltage Power Supply (LVPS) board. 

The instrument control electronics uses a radiation hard $\mu$-processor (e.g., Intersil HS-80C85RH) to perform the following functions: (i) receive and process the serial digitized data from the thermopile channels as well as provide a master clock and tagged encoded commands to the ASIC command decoders via a single line; (ii) mathematical operations on the science data such as averaging or offset corrections; (iii) data reduction, packetization, and routing of the science and housekeeping data to the Probe via a RS422 protocol; (iv) receive and act upon commands received from the Probe, e.g., active channel selection, setting temperature levels, or motor positions; (v) control stepper motor positions as well as decode their respective positions; (vi) provide stable temperature control to the instrument; and (vii) collect all temperatures and supply and reference voltages to form housekeeping/time stamped header packets that are streamlined into the data output to the Probe. All timing functions are synchronized with a 1 pulse per second (PPS) square wave from the Probe. The LVPS board accommodates DC-DC convertors and other various voltage/current control devices. This board not only conditions and regulates the voltages for various electronic usage but also controls power to the heaters. The Probe +28 VDC bus voltage is filtered and dropped via DC-DC switch mode converter to two main voltages: +3.3 VDC for logic use and +5 VDC for the stepper motor.

\subsection{Hera Probe Nephelometer}
\label{sec:3.4}

\subsubsection{Investigation Overview}
\label{sec:3.4.1}

Knowing the micro- and macro-physical properties of the haze and cloud particles in Saturn's atmosphere is crucial for understanding the chemical, thermo-dynamical and radiative processes that take place. Full characterization of the various types of haze and cloud particles requires {\it in situ} instrumentation, because Saturn's stratospheric hazes obscure the lower atmosphere, and because remote-sensing measurements of (for example) reflected sunlight depend on myriads of atmospheric parameters thus prohibiting reaching unique solutions. The {\it Hera} Nephelometer (NEPH) will illuminate haze and cloud particles, and will measure the flux and degree of linear polarization of the light that is scattered in a number of directions. The particle properties can be derived from the dependence of the scattered flux and polarization on both the scattering angle and the wavelength. 

The primary measurement objective of the Nephelometer is to characterize the micro- and macro-physical properties of atmospheric particles by measuring the flux and polarization of light that is scattered by particles that are passively sampled along the probe's descent trajectory. The angular and spectral distribution of the flux and polarization of the scattered light provides the particles' size distribution, composition, and shape, as well as their number density. NEPH's secondary objective is to measure the flux and polarization of diffuse sunlight in the atmosphere. This will provide the optical depth along the trajectory and its spectral variation, placing the results on the samples into a broad perspective. NEPH consists of two modules: Light Optical Aerosol Counter(LOAC) to measure the size distribution of particles, and Polarimetric Aerosol Versatile Observatory (PAVO) to also measure their shape and composition. The modules will be placed side by side to sample similar particles. 

LOAC's design is based on an instrument already in use as balloon payload for aerosol size determination in the Earth's atmosphere \cite{Renard15b,Renard15a,Renard10}. PAVO's optical design is based on the SPEX instrument \cite{Rietjens15} that is used on the ground to measure aerosol properties. The SPEX-optics has been tested successfully for radiation hardness with view of ESA's {\it JUICE} mission. A design similar to PAVO's (except for the polarimetric optical heads) is the nephelometer on the {\it Galileo} probe \cite{Ragent92}. Combining NEPH's results with ASI's ambient pressure measurements, the absolute vertical profile of the hazes and clouds along the probe's descent trajectory can be determined.

\subsubsection{Measurement Principles}
\label{sec:3.4.2}

The probe's descent through the atmosphere allows LOAC and PAVO to sample particles passively. The low solar flux levels in Saturn's atmosphere require both LOAC and PAVO to use artificial light sources for illuminating their samples.

\begin{figure}
\begin{center}
\includegraphics[scale=1.5]{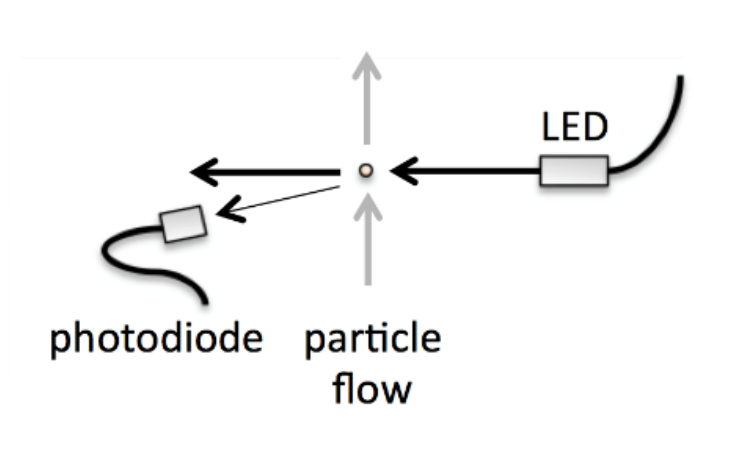}
\caption{Side-view of LOAC's design, with the particles crossing the LED light beam from below. The photodiode at a scattering angle $\Theta$~=~12$\degree$ captures the forward scattered light.}
\label{LOAC}      
\end{center}
\end{figure}

Figure $\ref{LOAC}$ shows a schematic of LOAC. Sampled particles cross a 2-mm diameter LED light-beam and the flux $F$ of the light that is scattered by a single particle across angle $\Theta$ = 12$\degree$ is measured. The scattered flux $F$ is very sensitive to the particle size, but relatively insensitive to its shape and/or composition. LOAC can accurately retrieve particle sizes between 0.1 and 250 $\mu$m in 20 size classes.

\begin{figure}
\begin{center}
\includegraphics[scale=1.4]{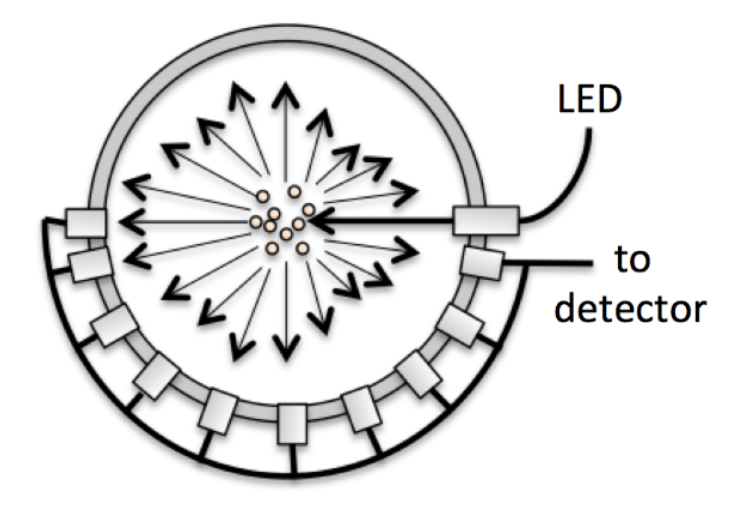}
\caption{Top-view of PAVO. One optical head captures non-scattered LED-light, and 9 heads capture scattered light. Fibres lead the modulated flux spectra from each head to the detector.}
\label{PAVO}      
\end{center}
\end{figure}

Figure $\ref{PAVO}$ shows PAVO's design. PAVO measures flux $F$ as well as degree $P$ and angle $\chi$ of linear polarization \cite{Hansen74} of light that is scattered by sampled particles at 9 angles $\Theta$:  12$\degree$ (the same as for LOAC), 30$\degree$, 50$\degree$ 70$\degree$, 90$\degree$, 110$\degree$, 130$\degree$, 150$\degree$, 170$\degree$. At each $\Theta$, a small optical head (without moving elements) translates the scattered light into two modulated flux spectra $F_M$: \\

\noindent $F_M$ ($\Theta$,$\lambda$) = 0.5 F($\Theta$,$\lambda$) [ 1 $\pm$ $P$($\Theta$,$\lambda$) cos $\Psi$],\\

\noindent where $\Psi$($\Theta$,$\lambda$) = 2$\chi$($\Theta$,$\lambda$) + 2$\pi$ $\delta$/$\lambda$, with $\lambda$ the wavelength and $\delta$ the delay of the optical retarder in the head \cite{Snik09} (see also Keller and Snik, patent application WO2014/111487 A1). The $\pm$ sign in the equation represents the beam-splitter in each head that produces two modulated flux spectra at every $\Theta$ that are subsequently fed to the spectrograph and detector with an optical fiber. Each modulated spectrum provides $P$ and $\chi$, while the sum of two spectra yields $F$. PAVO uses LEDs covering 400 to 800 nm to illuminate its sampled particles. 

An extra optical head at $\Theta$ = 0$\degree$ monitors variations of non-scattered LED-light to obtain information on the number of particles. By chopping the incident beam, we will be able to derive the local diffuse solar radiation field. Another, outward pointing optical head could be added to directly measure the diffuse solar flux and its polarization state.  

From the modulated spectra $F_M$ the scattered fluxes $F$ can be derived with a few nm resolution, and $P$ and $\chi$ with 10--20 nm resolution. The required accuracy for $P$ is 0.005 (0.5\%), well within the modulation technique's accuracy \cite{Rietjens15}.

\subsection{Hera Probe Radio Science Experiment}
\label{sec:3.5}

\subsubsection{Investigation Overview}
\label{sec:3.5.1}

The {\it Hera} Probe Radio Science Experiment will comprise two main elements. Radio tracking of the {\it Hera} probe from the Carrier Relay Spacecraft (CRSC) while {\it Hera} is under parachute will utilize the resulting Doppler shift to provide a vertical profile of zonal winds along the descent path for the duration of the probe telecommunications link detectability to at least ten bars \cite{Atkinson97,Atkinson96,Bird05}. The possibility for a measurement of the second horizontal component of the winds via a probe signal frequency measurement on Earth when the probe descends on the sub-solar side of Saturn \cite{Folkner06,Folkner97} will be carefully explored. The Radio Science/Doppler Wind Experiment (DWE) utilizes the {\it Hera} radio subsystem, knowledge of the {\it Hera} probe descent location, speed, altitude profile, and the CRSC trajectory to make a precise determination of the probe speed relative to the CRSC from which the zonal wind drift can be extracted. Additionally, as the {\it Hera} probe is expected to drift by up to several degrees in longitude under the influence of the zonal winds, the reconstruction of the probe descent location will provide an improved geographical context for other probe science investigations.

Additionally, the Radio Science/Atmospheric Absorption Experiment (AAbs) will utilize the {\it Hera} radio subsystem mounted on the CRSC to monitor the signal strength of the probe signal, providing a measurement of the integrated atmospheric absorption along the signal propagation path. The {\it Galileo} probe used this technique at Jupiter to strongly constrain the atmospheric NH$_3$ profile, complementing the atmospheric composition measurements of the probe Mass Spectrometer \cite{Folkner98}.

The primary objectives of the {\it Hera} Probe Radio Science Experiment are therefore to 1) use the probe radio subsystem (both mounted on the probe and the CRSC) to measure the altitude profile of zonal winds along the probe descent path with an accuracy of better than 1.0 m/s, and 2) to measure the integrated profile of atmospheric absorption, expected to be primarily due to NH$_3$ between the probe and CRSC. Secondary objectives include the analysis of Doppler modulations and frequency residuals to detect, locate, and characterize regions of atmospheric turbulence, convection, wind shear, and to provide evidence for and characterize atmospheric waves, and from the signal strength measurements, to study the effect of refractive-index fluctuations in Saturn's atmosphere including scintillations and atmospheric turbulence. \cite{Atkinson98,Folkner98}.

\subsubsection{Measurement Principle}
\label{sec:3.5.2}
 
The {\it Hera} Transmitter UltraStable Oscillator (TUSO) will generate a stable signal for the probe radio link. The Receiver USO (RUSO) will provide very accurate measurements of the probe link frequency. Knowledge of the probe descent speed and the CRSC trajectory will allow the retrieval of Doppler residuals due to unresolved probe motions including wind. While in terminal descent beneath the parachute, the vertical resolution of the zonal wind measurements will depend upon the probe descent speed \cite{Atkinson89}. In the upper atmosphere the vertical resolution will be on the order of 7 km, while in the deeper atmosphere variations with a vertical scale size on the order of one km can be detected. The accuracy of the wind measurement will primarily be limited by the stabilities of the TUSO and RUSO, the reconstruction accuracy of the probe and CRSC trajectories, and the relative geometry of the {\it Hera} and CRSC spacecraft. Assuming a UHF link frequency, a wind measurement accuracy better than 0.2 m/s is expected \cite{Atkinson98,Bird97}.

\subsubsection{Design Description / Operation Principle}
\label{sec:3.5.3}

The {\it Hera} probe telecommunications system will consist of the radio transmitter subsystem on the probe and the receiver subsystem on the CRSC. The carrier receiver will be capable of measuring the {\it Hera} telemetry frequency at a sampling rate of at least 10 samples per second with a frequency measurement accuracy of 0.1 Hz. The signal strength will be measured with a sample rate of 20 Hz and a signal strength resolution of .01 dBm \cite{Folkner98}. These sampling rates will enable probe microdynamics such as probe spin and pendulum motion, atmospheric waves, aerodynamic buffeting and atmospheric turbulence at the probe location to be detected and measured. 

The long-period stability of both the TUSO and RUSO, defined in terms of 30-minute fractional frequency drift, should be no greater than $\Delta$$f$/$f$ =10$^{-11}$, with an Allan Deviation (at 100-s integration time) of $\sim$10$^{-13}$.  The USO drift during the probe descent (90 minutes maximum) is 0.01 Hz.

\subsection{Hera Carrier Camera}
\label{sec:3.6}

\subsubsection{Measurement Objectives}
\label{sec:3.6.1}

The {\it Hera} carrier spacecraft will feature a simple visible camera, designed by the Laboratoire d'Astrophysique de Marseille with hardware contributions from the UK and Spain.  Design heritage is based on the $Rosetta$/OSIRIS camera \cite{Keller07}.  The purpose of this camera is fourfold:

\begin{itemize}[label=--]
\item To provide optical navigation;
\item To provide contextual imaging of the probe entry site, characterizing the morphology of nearby discrete cloud and haze features, waves and determine local cloud motions with a precision of 3 m/s; 
\item	To provide a global characterization of Saturn's atmosphere over multiple days during the approach of the carrier; 
\item	To provide contextual imaging of the Saturn system during the approach and departure phase, imaging the planet, rings and diverse satellites.
\end{itemize}

The camera will accomplish these three requirements with eight carefully selected filters mounted on a filter wheel (see Table \ref{filters}). In addition, the camera will be also used during the cruise to measure Saturn's brightness continuously and determine the planet's acoustic oscillation modes.

\subsubsection{Design Description}
\label{sec:3.6.2}
 
\begin{figure}
\begin{center}
\includegraphics[scale=1.4]{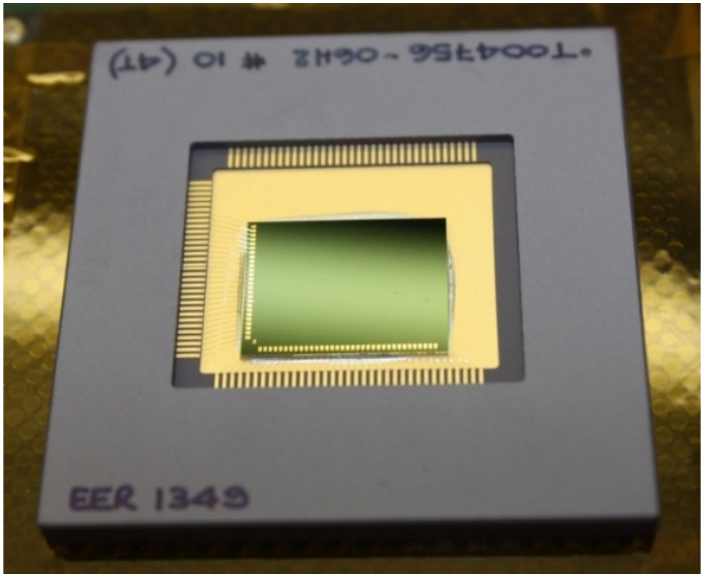}
\caption{CIS115 from e2v.}
\label{CIS}      
\end{center}
\end{figure}

The design of the camera is dictated by the compromise between the release distance of the probe from Saturn and the angular size of the planet at this distance. A good balance is found with a focal length of 160 mm, which, associated, with the pixel size of the selected detector leads to a resolution of 8.75 arcsec/pixel, when the field of view is close to 3.7$\times$5.0 square degrees. This compromise will be further studied during Phase A. The optical design is based on a dioptric design using radiation hard glasses, with a pupil diameter of 40 mm, (f/4), allowing exposure times in the 0.1-10 s range in spectral ranges selected by a 8 position filter wheel. The Complementary Metal Oxide Semiconductor (CMOS) active pixel sensor CIS115 from e2v (see Fig. \ref{CIS}) is preferred to a CCD mainly for its better radiation hardness and for its smaller pixel size, valuable characteristics for the compactness of the design. Moreover, this detector is Òpre-qualifiedÓ for another ESA space mission-{\it JUICE} \cite{Soman14}. Its size of 1504 $\times$ 2000 pixels provides the aforementioned field of view. The detector will be passively cooled to around -40$\degree$C to suppress radiation-induced bright pixels. The main structure of the camera is based on a tube made from aluminum alloy or from a similar thermally stable material, depending on the expected environmental conditions of the camera. The camera structure also supports the filter wheel mechanism, and the focal plane assembly, trimmed in position by use of adjustable shims. Three bipods support this structure and limit the possible mechanical stress coming from the bench, on which the instrument is mounted. The electronics box includes the control and command of the detector and of the filter wheel and possibly an automatic control of the exposure time to optimize the dynamics of the current image by analysis of the previous. Compression of the images, in real time, with adjustable compression rates including lossless compressions could also be added, depending on the possible data transmission volume. Packetized data are transmitted to the mass memory of the spacecraft in real time, as the instrument does not include a significant memory. Communications with the spacecraft are based on 2 redundant space wire links, for commands, image data and housekeeping. Without inner memory and without image compression, the image cadence is limited by the rate of this link (1 image per 2.2 s at a rate of 100 Mbits/s). Dual redundant DC-DC converters supply the various sub-systems with appropriate voltages. The electronics is based on the use of one processor (and a second in cold redundancy) ensuring all this task.

\section{Proposed mission configuration and profile}
\label{sec:4}

\subsection{Science Mission Profile}
\label{sec:4.1}

The Saturn probe science mission is a relatively short phase at the end of a relatively long transfer from Earth to Saturn (but shorter than $Rosetta$'s transfer). To provide context for the science mission this section first gives an ordered timeline of all the flight phases of the mission, followed by more detailed discussions of the phases central to delivering the science results.

\subsubsection{End-To-End Mission Profile}
\label{sec:4.1.1}

The Saturn probe mission begins its flight phases with a Soyuz-Fregat launch from Kourou on a transfer trajectory to Saturn. Though a thorough search for trajectory options has not yet occurred (it is planned for the assessment phase), multiple options for M-class missions are expected in future launch windows in the next decade. For study purposes the {\it Hera} team has used an example trajectory launching in May 2025, with Venus and Earth gravity assists in August 2025 and July 2026, respectively, arriving at Saturn in August 2033. Variants to this trajectory exist that shorten the trip time by as much as a year, at the expense of larger $\Delta$V requirements. Later launch windows will be studied. No science data acquisition is planned until Saturn approach, so operations for the great majority of the transfer are limited to activities such as trajectory and spacecraft systems monitoring and maintenance. During this period, a system of solar panels and secondary batteries provides electric power. Several weeks priort to Saturn arrival, the CRSC will turn to the proper probe release attitude and will release the {\it Hera} probe onto the probe's delivery trajectory, spinning for attitude stability. The probe continues on a ballistic trajectory until entry into Saturn's atmosphere. After probe release the CRSC performs more navigation observations and then a divert maneuver, placing it on a trajectory that allows the CRSC to be properly positioned for the probe data relay. The timing of the probe's release is a trade to be performed in the assessment phase.

The probe entry and descent sequence begins a few hours before entry when the probe event timer begins the ``wake up'' process, warming the probe's instruments and support systems in preparation for data acquisition and return. Upon encountering the atmosphere, an aerodynamically stable aeroshell enclosing the probe's Descent Module (DM) will protect the DM from the extreme heat and dynamic forces of entry into Saturn's hydrogen-helium atmosphere at speeds between 26--30 km/s. By this time the CRSC has begun its $\sim$90-minute overflight of the entry site, aiming the high gain antenna to receive data transmitted from the probe. After the hypersonic deceleration phase is over, the probe's aeroshell is jettisoned, a main parachute opens, slowing the DM's descent in the atmosphere's upper, less dense regions, and the {\it in situ} science instruments begin acquiring their data. As the DM descends into denser atmosphere, at an altitude and via a method to be determined in future trade studies, the descent rate will be increased, allowing the DM to reach the required depth (plus margin) during the CRSC overflight window. The DM transmits science data to the CRSC for as long as the probe-CRSC relay link survives but to at least the 10 bar pressure level and likely to the 20-bar design margin level or deeper. Eventually the combination of increasing pressures and temperatures will cause the DM systems to fail, then to melt, and finally to vaporize as the DM becomes a new part of Saturn's atmosphere. During the $\sim$70--90 minute DM descent the overflying CRSC, operating now on power from primary batteries, maintains the data relay link with the DM, storing multiple copies of the probe's data in redundant onboard storage media for later downlink to Earth. After the data reception window ends the CRSC turns its High-Gain Antenna (HGA) to Earth and begins the downlink, sending each data set multiple times. The CRSC's Saturn flyby places it on a solar system escape trajectory for spacecraft disposal. 

\subsubsection{Core Science Mission Profile}
\label{sec:4.1.2}

The Saturn probe's primary science mission closely resembles that of the {\it Galileo} probe, and has many similarities to ESA's {\it Huygens} probe that successfully entered and descended through Titan's atmosphere. After the warm-up period, the probe begins acquiring science data when its accelerometers detect acceleration due to atmospheric drag. Until the aeroshell is jettisoned there is no data relay to the CRSC, so the time-tagged accelerometer data and possibly heat shield recession data, needed to reconstruct the vertical profile of atmospheric density, are stored in onboard memory on the DM. When the aeroshell is jettisoned the radio system begins transmitting data from the now-operating {\it in situ} instruments, along with the stored data from the entry and deceleration phase. There is no radio receiver on the DM, so there is no real-time commanding capability of the DM after release from the CRSC. The CRSC uses the DM's radio signal, whose carrier frequency is controlled by an onboard ultrastable oscillator (USO), to make Radio Science Doppler Wind Experiment measurements during the descent, providing a measure of the vertical profile of zonal winds at the descent location. Using a command sequence loaded before release from the CRSC, a simple controller on the DM runs a pre-programmed series of measurements by each instrument and routes the data for storage and transmission. The controller uses temperature and pressure data from the Atmospheric Structure Instrument (ASI) to guide instrument modes and observation timing, optimizing the data set for science objectives appropriate to the different atmospheric depths. When the DM reaches the 10-bar level in Saturn's atmosphere, the data return strategy has all probe science data successfully transmitted to the CRSC, satisfying mission success criteria. Subsequent data are returned as the pre-determined (and diminishing) relay data rate allows, according to the controller's priority protocol, until increasing temperatures and pressures cause the DM systems to fail.

Initial analyses indicate that with some Saturn approach trajectories, such as the 2025 Earth-Venus-Earth-Saturn (EVES) trajectory mentioned above, the probe's entry location will be on the sunlit and Earth-facing side of Saturn, providing 90 minutes or more of descent before crossing the evening terminator. This is very beneficial for two potential experiments. Sunlight intensity measurements by a visible-wavelength channel on the Net Flux Radiometer allow inferring the depth at which solar energy is deposited in Saturn's atmosphere, important in determining what drives Saturn's winds and the overall energy balance of the atmosphere. Receiving the DM's carrier frequency at Earth, possible only when the DM is on the Earth-facing side of Saturn, allows a second Doppler tracking measurement to be made at Earth. This second wind vector component will help separate the line-of-sight wind speed at the probe location into zonal, meridional, and vertical wind speeds.

\subsubsection{Saturn Atmospheric Entry}
\label{sec:4.1.3}

\begin{figure}
\begin{center}
\includegraphics[scale=1.4]{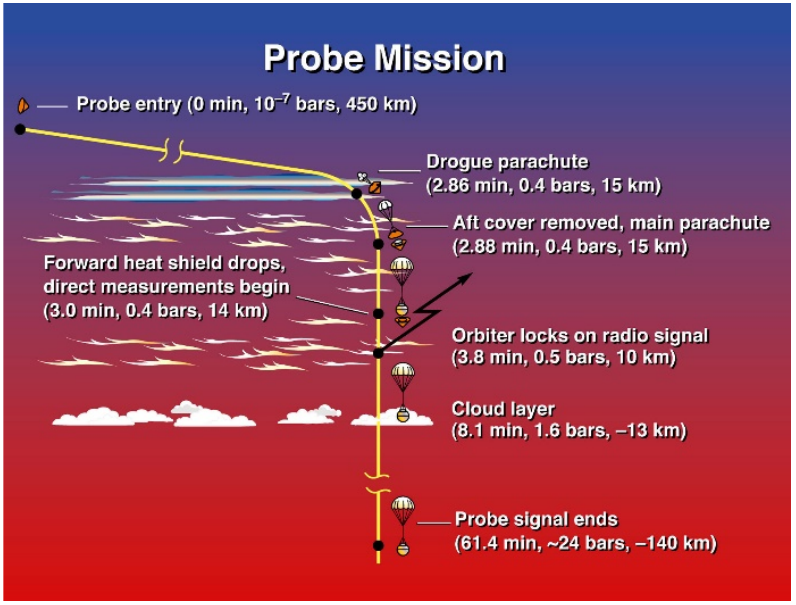}
\caption{Galileo entry, descent and deployment sequence shown above will be the basis for the proposed Saturn mission.}
\label{entry}      
\end{center}
\end{figure}

Entry into Saturn's atmosphere from hyperbolic approach is a difficult but manageable task. The proposed mission is similar to the {\it Galileo} Probe mission that entered Jupiter's atmosphere successfully in December, 1995, deployed the descent probe and collected and transmitted a wealth of data. The {\it Galileo} probe entered Jupiter's hydrogen-helium atmosphere at 47.4 km/s, compared to the 26--30 km/s range of entry speeds for the Saturn probe mentioned in Sec. \ref{sec:4.1.1}, resulting in a Saturn entry that is significantly less challenging than that faced by {\it Galileo} at Jupiter. Figure \ref{entry} shows the concept of operation for {\it Galileo} entry, deployment, and descent. The Saturn probe's entry and deceleration phase is very similar in most aspects to that of the {\it Galileo} mission. A probe scaled from {\it Galileo}'s 1.27 m diameter to 1.0 m, with an estimated entry mass of 200 kg as compared to {\it Galileo}'s 339 kg, can accomplish the required science at Saturn. Table \ref{Mass_estimate} uses the {\it Galileo} equipment as a basis for subsystem masses for the Saturn probe, and indicates that an entry mass of 200 kg is readily achievable. More rigorous design studies should allow significant reductions in structure mass, since inertial load levels will be much lower than {\it Galileo}'s design deceleration load of 350 $g$. Although the entry heating rate for a prograde Saturn entry is much less severe than {\it Galileo} experienced at Jupiter, Saturn's larger atmospheric scale height yields a long-duration entry resulting in a total heat load that is similar to the {\it Galileo} Jupiter entry. 

Ablative materials suitable for extreme entry missions and test facilities to qualify Thermal Protection System (TPS) for extreme environments are not yet available to ESA. Since the Heritage Carbon Phenolic (HCP) used for the {\it Galileo} and Pioneer-Venus missions is no longer available, NASA's innovative Heatshield for Extreme Entry Environment Technology (HEEET) now under development at NASA's Ames Research Center provides a very efficient solution for such an entry profile, resulting in a TPS mass that is only a fraction of the {\it Galileo} entry system's TPS mass. NASA plans for this technology to be available at TRL 6 by 2017 for mission teams currently proposing to its 2014 Discovery Program AO. In this context, HEEET is an appropriate technology for a Saturn probe mission that fits within the ESA M-class mission concept and satisfies the maturity requirements stated in the call. 

Entry velocity and Entry Flight Path Angle (EFPA) strongly influence the atmospheric entry challenge. Saturn's large planetary mass results in typical inertial entry speeds of 36 km/s or more, but during a prograde entry Saturn's high rotation rate mitigates up to 10 km/s of the entry speed, with the maximum benefit from a near-equatorial entry alignment. Steep EFPAs improve targeting accuracy and reduce the heat load but increase peak deceleration load, heating rate, and peak pressure. Mission success can easily be accomplished with an entry latitude below 10$\degree$ and EFPA between -8$\degree$ and -19$\degree$. Table \ref{g-load} summarizes the range of entry conditions and associated TPS mass for relevant combinations of EFPA and latitude. For all cases, the HEEET material is significantly more mass efficient than the HCP used for {\it Galileo}. The benefit is most pronounced at the shallower entry angles, which also provides more benign inertial loads. For steeper EFPAs, the ablative TPS mass is further reduced and is only 10\% of the entry mass. In the study that follows, we primarily focus on the EFPA =  -19$\degree$ case, corresponding to a probe entry system mass of 200 kg.

\begin{figure}
\begin{center}
\includegraphics[scale=1.4]{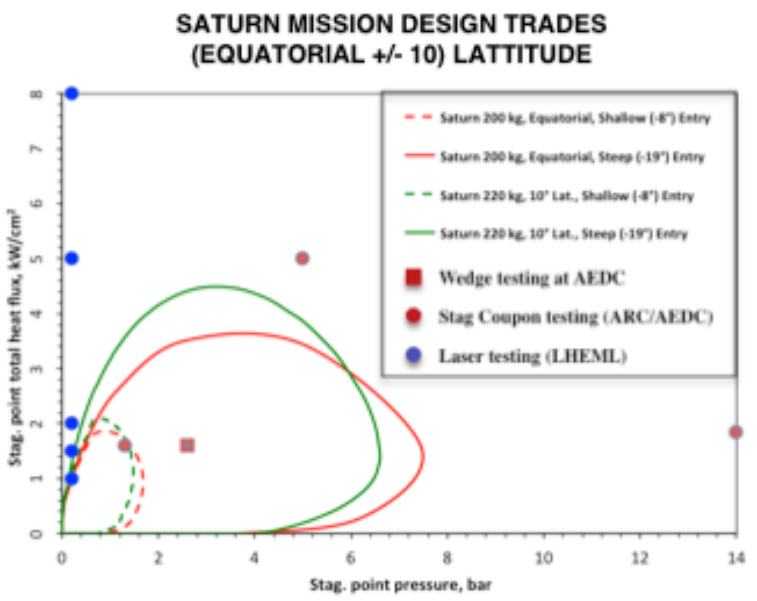}
\caption{Heat-flux and pressure (stagnation values) along four trajectories that bound the proposed Saturn mission is shown above along with arc-jet test conditions where HEEET has been tested. The HEEET acreage material shows exceptional performance with no failure even at extreme conditions (14 atmospheres and 2000 W/cm$^2$).}
\label{heat}      
\end{center}
\end{figure}

Figure \ref{heat} shows the stagnation point heat-flux and impact pressure along trajectories that are bounded by $\pm$10$\degree$ latitude (including equatorial) with EFPAs between -8$\degree$ and -19$\degree$. Also shown in this figure are the conditions at which HEEET material has been tested in arc-jet and laser heating facilities. HEEET acreage material is very well behaved at these extreme conditions and at shear levels that are far greater than the anticipated Saturn entry conditions. Adoption of HEEET will minimize the TPS technology risk for this mission.

\subsubsection{Probe Delivery to the Entry Trajectory}
\label{sec:4.1.4}

Since the entry probe carries no propulsive capability, it is on a ballistic trajectory from the moment of release and the CRSC must establish the proper entry probe trajectory and orientation upon release. The probe also has no active attitude control capability, so the CRSC must spin the probe to maintain its attitude until entry. After the long cruise from Earth to Saturn approach the first activities in preparation for release are navigation observations, leading to Trajectory Correction Maneuvers (TCMs) to establish the proper Saturn entry trajectory. The CRSC will release the probe at a distance from Saturn that ensures the entry trajectory will be within tolerances, and might image the departing probe to verify release accuracy and decrease the uncertainty in the probe location. Soon after probe release, the CRSC performs a divert maneuver, changing the CRSC trajectory to a Saturn flyby that provides data relay for the entry probe. The precise timing of probe release is a trade involving navigation accuracy, which degrades with increasing distance from Saturn (earlier release and longer probe coast), the mass of batteries needed to keep the probe warm during its post-release coast, and the mass of propellant needed for the CRSC's divert maneuver, which increases with decreasing distance to Saturn (later release and shorter probe coast). Assessment phase studies will estimate the optimum timing of those first navigation activities and TCMs, and probe release.

\subsubsection{Data Relay}
\label{sec:4.1.5}

The mission data return strategy uses the data relay method. Studies have shown that this approach yields higher data rates with less operations risk than the Direct-To-Earth (DTE) approach and carries other science benefits as well. Similar to the {\it Galileo} probe, after deploying from the entry aeroshell, the descent module transmits data over two independent channels (left and right circular polarization at slightly offset frequencies) through a ultra-high frequency (UHF) patch Low-Gain Antenna (LGA) on the DM's upper surface. The CRSC trajectory is within the LGA beam from the start of the probe data transmission through the end of the descent module's mission, some 70-90 minutes later. The CRSC points its HGA, feeding a UHF receiver, at the probe entry site, receiving both channels of probe data and storing them onboard in multiple redundant copies. Extremely conservative link analyses based on an 8 W UHF transmitter suggest data rates of at least 500 bps per channel (the {\it Galileo} probe data rate was 128 bps per channel). More refined analysis indicates a variable data rate is feasible, with rates potentially greater than 10 kbps for part of the descent. Performance far greater than the {\it Galileo} probe performance is enabled largely by two differences from the {\it Galileo} mission: Saturn's lack of intense radiation belts and their associated RF synchrotron radiation noise allows using UHF, which is less attenuated by atmospheric ammonia and water; and the distance from the DM to the CRSC during data relay is between 50,000 and 70,000 km, much closer than the 200,000+ km range of the {\it Galileo} relay.

After receiving all the probe data onboard, the CRSC downlinks the data to Earth via standard ESA communications facilities. The CRSC will turn its HGA to Earth, transmitting multiple copies of each redundant data file at X-band until the CRSC primary battery charge is effectively exhausted. After recharging its secondary batteries, it then repeats those transmissions periodically as the battery charge allows, until ground commands verify the full data set has been successfully received. Any ancillary data, such as context imaging from a CRSC imager, are included in this downlinked data set.

\subsection{System Level Requirements}
\label{sec:4.2}

\subsubsection{Entry Probe Requirements}
\label{sec:4.2.1}

Between release from the CRSC and atmospheric entry there are three primary requirements on the entry probe: 1) maintain orientation for entry; 2) maintain the probe subsystems and instruments within their environmental tolerances to ensure proper operation during entry and descent; and 3) provide adequate timing so the ``wake-up'' sequence begins at the proper time. The CRSC orients the probe and spins it prior to release for attitude stability, so the first requirement becomes a requirement on the probe's mass properties: its principle inertial axis must co-align with the aeroshell's symmetry axis. Maintaining environmental conditions is primarily keeping the DM warm at more than 9 AU. In the absence of radioisotope heater units, this will likely require primary batteries to power electric heaters. Also, new electronics technologies may provide electronics circuits that can operate at Cryo temperatures without the need for heaters. Batteries would also power an event timer of sufficient accuracy that the wake-up sequence is initiated in time to be completed before entry begins, but not so far in advance that it wastes battery power waiting for entry.

Atmospheric entry involves a different set of requirements. There is a new constraint on the probe's mass properties, along with its exterior geometry: the entire entry system (DM + aeroshell) must be aerodynamically stable at hypersonic speeds, and must maintain that stability in the face of ablative mass and geometry changes. The system must accommodate the extreme heating environment and potentially large inertial loads of atmospheric entry at 26--30 km/s. If the heat shield is instrumented, the heat shield sensor data must be stored onboard until the sensor and entry accelerometry data can be telemetered to the CRSC after DM deployment from the heat shield. After the end of the hypersonic deceleration phase the DM's controller must initiate a sequence of deployments, including drogue parachute mortar firing, backshell release and main parachute deployment, and heat shield release, for transition to the stable descent phase and primary science data acquisition.

When the DM stabilizes under its main parachute its controller must initiate radio transmission of data to the CRSC, and operation of all science instruments. This continues to a depth determined by the controller using ASI data, when the descent rate must be increased to reach the required 10-bar depth before the CRSC's received signal falls below a margined SNR limit. Potential methods include releasing the main parachute and freefalling or opening a smaller parachute (as per {\it Huygens}), reefing the main parachute, or other options, all to be studied in the assessment phase. During this descent phase the DM must maintain its systems within their operating environmental ranges while exposed to exterior temperatures and pressures from $\sim$85 K at 0.1 bar (near the tropopause) to $\sim$300 K at 10 bars, possibly increasing to $\sim$350 K at 20 bars.

Despite its fundamental nature and extreme importance to planetary science the data volume for the {\it Galileo} Probe mission was quite small, less than 1 Mbit. The threshold Saturn probe mission data volume will be of similar size. Studies suggest that data rates for the DM-to-CRSC link might support data volumes as high as several tens of Mbit, providing capacity for ancillary science investigations while retaining large margins. Because data acquisition modes change with depth, the data rates from DM instruments are expected to vary during the descent, so some data will need to be stored on the DM prior to transmission. The size of onboard memory required will be studied in the assessment phase, but certainly will be less than the size of the entire data set, and devices with tens of Mbit capacity are small and require little power.

\subsubsection{Carrier-Relay Spacecraft Requirements}
\label{sec:4.2.2}

The Saturn probe mission's CRSC is a fully capable spacecraft that supports the entry probe's mission with a wide variety of functions during cruise, Saturn approach, and the science mission. Preliminary studies resulted in the CRSC requirements discussed here, and produced a system-level spacecraft design and mass budget estimates for assessing feasibility, discussed in Sec. \ref{sec:4.4}.

During the Earth-to-Saturn cruise the CRSC provides all functions for delivering the combined spacecraft (probe + CRSC) to Saturn approach, and for maintaining the proper function of its own and the probe's systems, including environmental control, power, and data communications to and from the probe for periodic checkouts and post-launch entry sequence loads. Except for brief periods for activities such as TCMs, the solar-powered CRSC must point its solar arrays sunward, with relatively loose pointing requirements. Communications while in the inner solar system must accommodate uplink for commanding, navigation, and downlink of engineering data over a fairly wide range of Earth-spacecraft-sun angles. Cruise at heliocentric distances greater than 5 AU places more emphasis on power generation and communications. Normal communications will use a 1.5 m HGA, whose X-band beam width of $\sim$0.75$\degree$  half-width half maximum (HWHM) sets the spacecraft's pointing accuracy requirement of 0.25$\degree$.

Upon Saturn approach the CRSC delivers the probe to its entry trajectory at the proper attitude, and then diverts to a trajectory that allows it to provide data relay support. Required accuracies for navigation, trajectory control, and release attitude control will be studied in the assessment phase, but will not strain current technologies. During the probe's descent the CRSC must continually point its HGA to the entry/descent location to receive the probe's UHF data relay signal. At UHF frequencies the HGA beam width is wide, $\sim$16$\degree$ HWHM. Multiple copies of the entire data set must be stored in the CRSC before it downlinks the data to Earth, but those requirements are also easily met with a few hundred Mbit of storage capacity. After the data reception period is over, the CRSC must repoint the HGA to Earth, switch back to X-band, and downlink the data to ground stations. The CRSC must be capable of downlinking each copy of the data set to Earth ground stations at least twice to ensure transfer of the entire set. 

In all post-launch phases the CRSC handles all propulsive maneuvers. The post-release divert maneuver, and a deep space maneuver before the Earth flyby, are the mission's only deterministic maneuvers. The size of the divert maneuver depends upon its timing: the farther from Saturn, the lower the $\Delta$V required. At $\sim$30 days before probe entry (that $\Delta$V is $\sim$50 m/s (80 m/s budgeted); 15 days out it is nearly 100 m/s. Conservative estimates of statistical $\Delta$V budgets and margins indicate 315 m/s of $\Delta$V capability is sufficient.

\subsubsection{Ground System Requirements}
\label{sec:4.2.3}

The Saturn probe mission uses only standard ground system facilities and resources. The operations team will need standard office, computing, and communications facilities, and access to a mission control facility. Spacecraft commanding and engineering and science data downlink will use standard deep space communications facilities operating at X-band. High-activity periods will include launch, planetary gravity assist flybys, preparations for and execution of probe release and CRSC divert, and the science mission and subsequent data downlink. 

\subsection{Launch and Transfer Trajectory}
\label{sec:4.3}

\begin{figure}
\begin{center}
\includegraphics[scale=1.4]{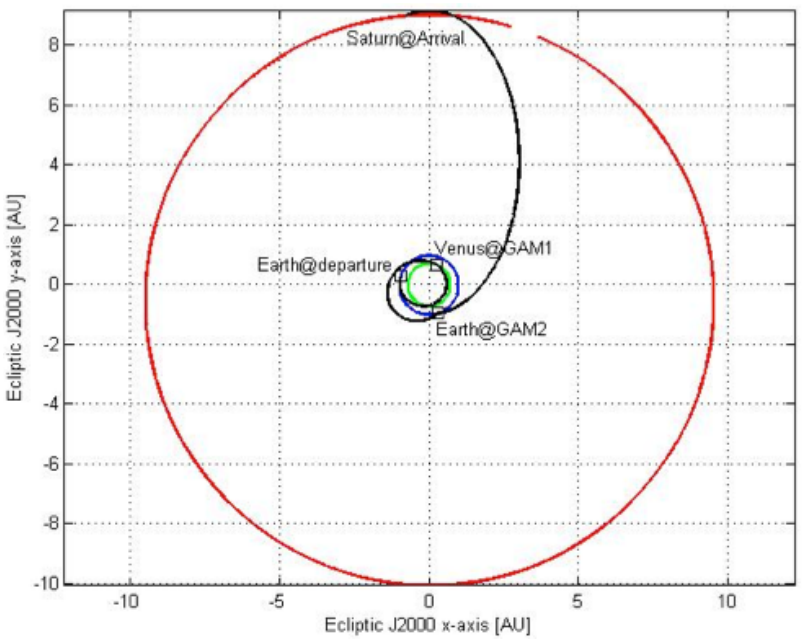}
\caption{Example of possible trajectory for Hera.}
\label{Traj}      
\end{center}
\end{figure}

One of the Saturn probe mission's greatest challenges is sending the spacecraft from Earth to Saturn. Although a thorough study of Earth-to-Saturn trajectories in time frames appropriate to ESA M-class flight opportunities has not been done, studies to date suggest there are multiple opportunities and trajectory types to consider. A wider search for Earth-to-Saturn trajectories will be an assessment phase task. The proposal team has used an example trajectory, identified as ``EVES 2025'' launching in May 2025 with gravity assists at Venus in August of 2025 and Earth in October of 2026, reaching Saturn in October of 2033 (see Figure \ref{Traj}). Launch is to a departure $V_\infty$ of 4 km/s (C3 of 16 km$^2$/s$^2$), sufficient to reach Venus, at a declination of 2.76$\degree$. The Soyuz-Fregat launch vehicle has a capacity of 1400 kg to this $V_\infty$. There is a Deep Space Maneuver (DSM) between the Venus and Earth flybys with a $\Delta$V of 190 m/s. Statistical $\Delta$V includes a launch residuals cancellation maneuver soon after launch, budgeted at 10 m/s, and Trajectory Correction Maneuvers (TCMs), budgeted at 20 m/s. Saturn arrival is a few months after the spacecraft's aphelion, at a $V_\infty$ of 6.203 km/s and approach declination of only 0.077$\degree$ (i.e., a nearly equatorial approach). This approach yields a probe entry location well into the sunlit and Earth-facing side of Saturn. Entry at a near-equatorial latitude and at a -14$\degree$ entry flight path angle (EFPA) would be at an atmosphere-relative speed of slightly more than 26 km/s, near the theoretical minimum for a hyperbolic approach from an Earth-to-Saturn transfer. Entries at higher latitudes, and at steeper EFPAs, would increase that speed somewhat, as discussed in Sec. \ref{sec:4.1.3} above. {30 days before the fly-by (value assumed for the study)}, the spacecraft is awakened from hibernation, the CRSC trajectory is adjusted for probe entry targeting and the probe is released. An avoidance manoeuvre (divert manoeuvre) is then performed to provide for the adequate fly-by altitude. The spacecraft then goes back to hibernation to save primary batteries mass and is awakened just prior to the probe mission so the CRSC can support the probe communication relay.

\subsection{Flight System}
\label{sec:4.4}

\subsubsection{Entry Probe}
\label{sec:4.4.1}

The entry probe element consists of two major sub-elements: the DM that carries all the science instruments and support equipment; and the aeroshell that protects the DM during transfer cruise, post-release cruise, and atmospheric entry, keeping the DM safe from pre-launch until the hypersonic deceleration phase is finished. A 200 kg probe mass estimate is based on dimensional scaling laws applied to the {\it Galileo} probe and first-order adjustments for different instruments and use of the HEEET TPS materials discussed in Sec. \ref{sec:4.1.3} above. No adjustments have been made for the Saturn probe's more benign entry conditions, e.g. lower inertial load, so more detailed study during the assessment phase might realize further mass savings.

\paragraph{\it Descent Module}
The Descent Module has four primary functions:

\begin{itemize}
\item House, control, provide power to, and maintain the operating environment of, science instruments and DM subsystems;
\item Collect, store (as needed), and transmit to the CRSC all science and engineering data;
\item Control the descent rate profile of the DM to satisfy science objectives and operations requirements;
\item Initiate the "wake up" sequence at the proper time before atmospheric entry.
\end{itemize}

The DM must survive the post-release cruise and the atmospheric entry. Surviving cruise is mostly a matter of electric power for small heaters, along with thermal insulation on the aeroshell exterior. The probe's primary battery complement is sized to include that function. Use of European Radioisotope Heaters Units (RHUs) would significantly decrease that battery size, but they are not used in this preliminary design due to low TRL. Surviving atmospheric entry involves robustness to large inertial loads of tens to possibly 100 $g$'s or more. The DM relies on the aeroshell for protection against the intense heating and huge thermal loads of entry.

All functions except descent rate control would use {\it Galileo} techniques. Once it deployed its main parachute the {\it Galileo} probe did nothing to control its descent rate. Like the {\it Huygens} probe, the Saturn probe cannot afford that simplicity because staying on the unmodified main parachute for the entire descent results in an excessively long descent duration making it impossible to reach 10 bars in the time available for the probe data link. During the assessment phase the DM's descent rate profile and several candidate approaches for controlling the DM's descent rate will be examined. The primary battery approach for the power source is retained, but batteries are now available with higher specific energy resulting in potential mass savings.

\paragraph{\it Aeroshell}

NASA and NASA Ames Research Center would provide the entire aeroshell, although alternatives from within the European community will be explored. The aeroshell consists of two main segments, a foreshell and a backshell, and has five primary functions:

\begin{itemize}
\item Provide an airframe that is aerodynamically stable at hypersonic and supersonic speeds in an H$_2$-He atmosphere, and is spin-stable along its symmetry axis;
\item Protect the DM from the intense heating and huge thermal loads of entry;
\item During hypersonic entry, accommodate the large deceleration loads from the DM;
\item Provide a stable transition from supersonic to subsonic flight;
\item Upon completion of its entry functions, separate from the DM (by command from the DM).
\end{itemize}

Section \ref{sec:4.1.3} above treats the entry aspects in detail and discusses probe size and mass. The preliminary studies used an estimated foreshell diameter of 1 m and a total mass of 200 kg. The aeroshell's role in entry heating protection also gives it a role in post-release survival: its thermal insulating properties aid retention of heat in the DM. The shape of the aeroshell is important. The {\it Galileo} probe foreshell, a 45$\degree$ sphere-cone, provides heritage for stability and ability to handle the thermal environment. The much lighter backshell of the aeroshell must protect from convective heating by hot gas from the foreshell, and from radiative heating from the trailing shock, where the atmospheric gas ``blown open'' by the probe's passage ``slams shut'' again. A partial-sphere shape is appropriate for the backshell, with the entry probe's center of mass at the center of the sphere. With that alignment, odd pressure distributions on the backshell resulting from turbulence or atmospheric winds can cause translational movements but not the much more troublesome angular movements (i.e., rotation) that could destabilize the probe. Transition to subsonic flight overlaps with aeroshell deployment. Most entry aeroshell shapes are unstable during the transition from supersonic to subsonic flight and need stability enhancement. Having the backshell deploy a drogue parachute while at a Mach number comfortably above unity can provide the additional stability, a technique the {\it Galileo} probe used. After slowing to subsonic speeds the drogue parachute provides sufficient drag force to pull the backshell from the probe, and then to deploy the DM's main parachute. The main parachute's drag is sufficient to pull the DM from the foreshell.

\subsubsection{Carrier-Relay Spacecraft}
\label{sec:4.4.2}

Spacecraft traveling to the outer solar system face three main design challenges: power generation, telecommunications (``telecom''), and thermal control. The Saturn probe mission is a flyby mission, so unlike an orbital mission, propulsion is not a strong design driver. Flight system design work did not attempt to optimize the design, but instead provide a conservative proof-of-concept design with sufficient margins to accommodate trades and design work in subsequent phases. All trades done to date will be reopened in the assessment phase and some will likely continue after the assessment. The result of work to date is a system-level design that meets all the requirements and fits within M-class resource constraints and expected schedules. This section first describes the designs for the three most challenging systems, then describes the remaining systems and configuration.

\paragraph{\it Power Technology and Power Sizing}

With operations at up to 9.5AU from the sun, power generation technology was the main driving requirement of the preliminary design task. This mission would benefit greatly from nuclear power sources, whether for power generation or for heating (RHUs), but use of European launchers requires use of European nuclear technology for launch pad safety qualification. The cost and low TRL of European space nuclear technologies makes that option difficult to achieve in the next ten years, so they were not used in this preliminary study. Fortunately the mission's brief science operations phase requires relatively little energy. A solar power system supplemented by primary and secondary batteries is sufficient:

\begin{itemize}
\item	The need for survival heating at up to 9.5 AU and for an on-board clock sets the solar array size. A low-activity power demand of 140W requires a solar array output of 170W at 9.5AU, provided by a classical rigid solar array with Low Intensity Low Temperature (LILT) cells, with a surface area of about 53m$^2$;
\item Secondary batteries provide energy for safe mode during cruise and allow for temporary off-sun pointing before arrival at Saturn (for trajectory correction maneuvers (TCMs), probe spinup and release, etc.);
\item Primary batteries supplement the system during the science operations phase. A supplemental capacity of 16.6 kWh meets the 400W power demand during science operations (with telecom system transmission) for a 64h time period.
\end{itemize}

For conservatism in this preliminary study we used the highest-TRL solutions for the solar array and primary batteries:

\begin{itemize}
\item Standard rigid solar array using Spectrolab LILT cells (currently flying on {\it Juno});
\item Saft LSH20 primary batteries currently flying on the $Rosetta$ mission's Philae lander.
\end{itemize}

A delta-qualification will be required for the LILT cells (qualified in Jupiter conditions) to qualify them for Saturn conditions. In the current baseline, the Solar Array is provided by the US. To further expand qualification status to Saturn, Boeing Spectrolab's heritage next triple junction (29.2\% eff) solar cells can be screened and tested for operation under 9.5 AU LILT conditions using their X-25 solar simulator with specially tuned neutral density filters. Based on their experience on {\it Juno}, Spectrolab's assessment is that the probability of success is elevated and that the TRL of 5 can be acquired in parallel with the Assessment Phase before 2018.

The Solar Array and the primary batteries would however represent then more than one third of the CRSC dry mass. During the assessment phase, lower TRL advanced technology options can be traded so as to decrease the mass and/or provide more operational time, including:
\begin{itemize}
\item Flexible solar array design;
\item Concentrator-based solar array design;
\item {Higher performance primary batteries};
\item Decrease hibernation/survival heating need thanks to RHUs;
\item Cryo Electronics do not need heaters.
\end{itemize}

\paragraph{\it Telecommunications and Data Handling}

For communications from the CRSC to Earth the telecom system uses a 65W X-band Traveling Wave Tube Amplifiers (TWTA) through a 1.5 m 2-axis steerable HGA, providing 4--5 kbps downlink at 9.5 AU. LGAs ensure communication with Earth when the spacecraft is in Earth's vicinity, and low-rate communications in safe mode. Data reception from the probe also uses the HGA, feeding a UHF receiver. The mission's data volume is quite small, possibly several tens of Mbit or less\footnote{Camera science would be decoupled temporally from Probe science since the camera is on the CSRC.}, so depending on the choice of on-board computer, the computer's standard memory could store multiple copies of the probe data. If necessary, a small supplemental memory device can expand available memory to meet the storage requirement. Each copy of the probe data is downlinked multiple times for redundancy.

\paragraph{\it Thermal}

Given the large range of heliocentric distances for any Earth-to-Saturn trajectory, maintaining an acceptable thermal environment for spacecraft subsystems without excessive power requirements is a non-trivial task. Particular attention must be devoted to propulsion system propellant tanks and feed lines to externally mounted thrusters. Power needed to keep those components warm when more than 5 AU from the sun is a significant part of the solar array sizing budget. The $Rosetta$ mission successfully addressed slightly less demanding but nonetheless similar challenges. With careful attention during design to heat loss pathways and appropriate use of lightweight insulating materials, high-TRL materials and techniques will suffice for a robust thermal management system.

\paragraph{\it Propulsive Delta-V}
To reduce mass a Mixed Oxides of Nitrogen (MON) Monomethylhydrazine (MMH) bi-propellant subsystem is adopted for this preliminary study. No main engine is required: 10N thrusters are sufficient. Propellant temperature must be considered since propulsion system heating to prevent propellant freezing is a major factor in the survival-heating budget, which sizes the solar array.

Table \ref{DeltaV} shows the $\Delta$V and propellant budget for the trajectory presented in Sec. \ref{sec:4.3}, assuming a dry mass of 1115 kg including the 20\% system margin (875 kg dry after probe release), a propellant need of 9kg for GNC, times two to account for the 100\% required margin, and 2\% of residuals. 

\paragraph{\it Pointing Modes and AOCS Design}
The CRSC is designed to remain sun-pointed throughout cruise except for brief periods for activities such as TCMs and probe release. For any trajectory that stays beyond 0.7 AU the LILT solar cells are compatible with full sun exposure at Venus and do not require off-pointing when close to the Sun. Safe mode is also sun-pointed. The cruise mode is in slow spin, interrupted only for TCMs and Earth communications, which use the HGA's two-axis pointing mechanism so the solar arrays stay sun-pointed. The mission uses a hibernation mode similar to that of the $Rosetta$ mission. Beyond about 5.5 AU, the solar array output is insufficient to power continuous operation, requiring hibernation from 5.5 AU to Saturn approach. It can be used also inside of 5.5 AU to decrease operations costs. During hibernation the CRSC is sun-pointed, in slow spin. The solar array panel arrangement aids stability by establishing the maximum moment of inertia around the spin axis.

At the proper time before the fly-by (to be established in the assessment phase), the CRSC will wake up, send status data to Earth to verify proper spacecraft function, and perform navigation tasks to allow design and execution of a TCM, if needed. At the proper time it enters the inertial pointed mode, executes the TCM, turns to the attitude needed for probe release, releases the probe, turns to the divert maneuver attitude, performs the maneuver, and then returns to sun-pointed hibernation mode. Studies during the assessment phase will establish the optimum timing for this sequence.

For the data relay fly-by the CRSC will be sun-pointed in inertial pointing mode. It will wake up, point its HGA towards the Earth for a systems check, then towards the probe entry site for receiving the probe's data while it descends, then towards the Earth again for data downlink. If reaction wheels are used, duty cycles will be very limited and will allow use of very light wheels. Without reaction wheels, 1N ``fine'' thrusters may be used instead. This tradeoff will be investigated during the assessment phase.

\subsubsection{Probe Delivery}
\label{sec:4.4.3}

The time between probe release and entry will be investigated to optimize probe primary batteries mass and propellant mass required for the post-release divert maneuver. The shorter this time, the lighter the probe batteries, but the higher the required $\Delta$V for the CRSC divert maneuver. For these preliminary studies we assumed release 30 days before entry. Navigation and attitude control accuracy requirements will be investigated as well.

\subsubsection{Mass Budget}
\label{sec:4.4.4}

Table \ref{mass2} presents nominal mass estimates based on similar subsystems from previous missions. The values are given in maximum estimated mass (best estimate plus maturity margin) so they can be compared to similarly-sized past planetary missions. A 20\% system margin is added for a conservative estimate of propellant mass and wet mass. Adding the launch adapter (Soyuz 937mm diameter interface) yields the mass to be compared with the mass capacity from the trajectory analysis. The dry mass structural index used is comparable to same-sized planetary missions: this is conservative as the CRSC will have fewer components than an orbiter. The resulting mass is compatible with the Soyuz launch scenario presented in Section \ref{sec:4.3}.

\subsubsection{Configuration}
\label{sec:4.4.5}

A preliminary configuration design allowed checking for fundamental physical architecture issues. One suitable face allocation was found but many other combinations may prove of interest during future phases. We show here an example of a 6-faced physical architecture including:

\begin{itemize}
\item	The launcher interface face, also hosting the main thrusters; this face is sun-pointed in cruise;
\item Two lateral faces each supporting one solar array wing (27m$^2$ each);
\item One lateral face supporting the HGA during launch;
\item One lateral face for CRSC bus units;
\item One top face bearing the probe.
\end{itemize}

\begin{figure}
\begin{center}
\includegraphics[scale=1.4]{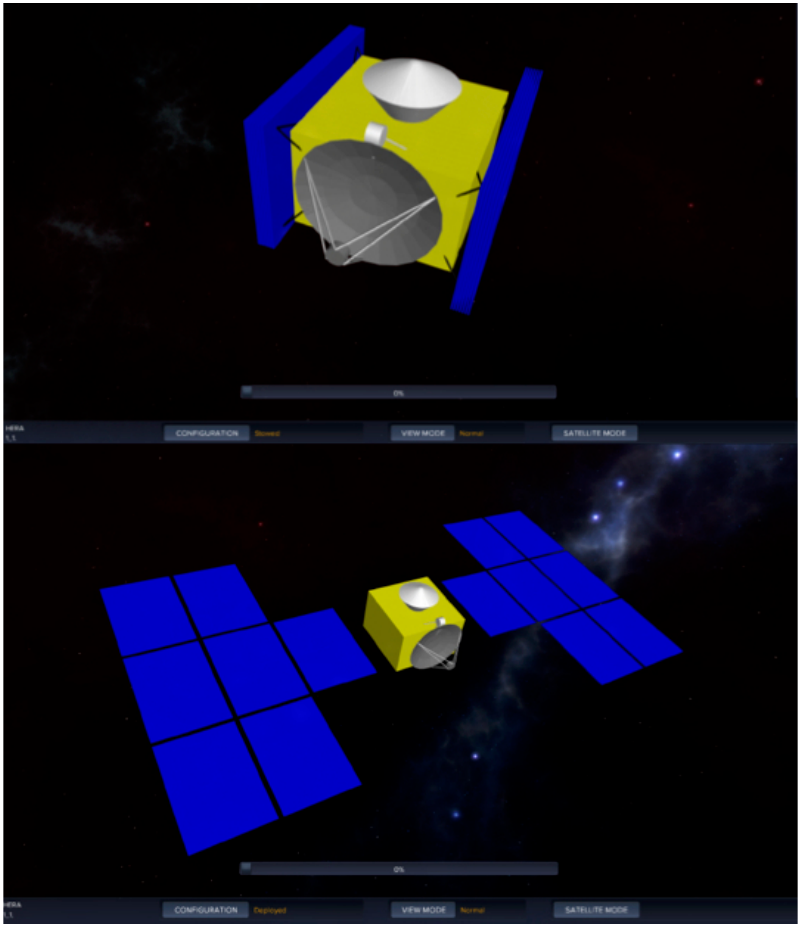}
\caption{7-panel solar array arrangement stowed (top) and deployed (bottom).}
\label{arrays}      
\end{center}
\end{figure}

The solar array is deployed shortly after launch so that its cells are towards the launcher interface face. The HGA is then deployed towards this face. Figure \ref{arrays} shows a 7-panel solar array arrangement stowed and deployed. This 1-3-3 arrangement has flown on telecommunication satellites like Ciel2 with bigger panels. It yields a large moment of inertia about the spin axis for robust spin stability.

\section{Management scheme}
\label{sec:5}

\subsection{Management Overview}
\label{sec:5.1}

The {\it Hera} mission for an {\it in situ} atmospheric probe of Saturn is proposed as an ESA-led mission, with a significant and essential contribution by NASA.  
Participating in the Proposal Consortium are two industrial companies, Thales Alenia Space France and The Boeing Company in Huntington Beach, California, USA.  Additionally, six science teams will participate in the Phase A Study, representing six potential science instruments for the {\it Hera} mission.  
If {\it Hera} is selected for flight following Phase A then a joint ESA-NASA mission management will be established under the responsibilities of both agencies. ESA and NASA will follow their own approach for the industrial activities. At the appropriate time during the study phase, ESA will select its industrial contractor for the study phase B1, (or contractors if parallel competitive studies are being conducted) and in a second step the {\it Hera} development industrial contractor (Phase B/C/D/E/F). 

\subsection{International Collaboration}
\label{sec:5.2}

The international collaboration for the {\it Hera} mission will involve the Spacecraft carrier bus and atmospheric probe as well as the science instruments and science investigations.  The Phase A study of spacecraft structures and systems will involve selected industrial partners, as well as NASA Ames Research Center and Jet Propulsion Laboratory. The selected European industrial partner will study the architecture for the Carrier Spacecraft Bus and the Probe. The selected American industrial partner will contribute by studying the solar power system, including solar arrays and the power management and distribution system. NASA Ames Research Center will contribute by studying the probe entry system, including the Thermal Protection System (TPS) for foreshell and backshell and the associated underlying structure including aero-thermodynamic.

The {\it Hera} instrument payload will be provided by instrument PI teams from ESA's Members states and NASA scientific communities. The different science instrument consortia are described below and summarized in Table \ref{table_consortia}. Payload funding for ESA's members states will be provided by National funding agencies. US payload contribution will be funded by NASA. The lead-funding agency for each PI-team will either be the PI National Funding Agency for a European PI-led team and NASA for a US-led PI team. NASA funding decisions will not be made prior to selection for the Phase A study.

\label{sec:5.4.3}

\subsection{Data Policy}
\label{sec:5.4.7}

The main repository for an ESA-led Planetary mission is ESA's Planetary Science Archive (PSA). Science instrument data will be archived in a timely manner in ESA's PSA, and the data will be mirrored to NASA's Planetary Data System (PDS). 

Our project plans to share these experiment data with the outer planet community through participation at symposia and workshops. Additionally, we will present papers and posters at relevant planetary science professional meetings and workshops, such as the European Geosciences Union (EGU), the American Geophysical Union (AGU), the European Planetary Science Congress (EPSC), the Lunar and Planetary Science Conference, NASA's Outer Planet Analysis Group (OPAG), and the Division for Planetary Science (DPS) Unit of American Astronomical Society. A special effort will be made to collaborate with the Exoplanet community in achieving a broader context for the {\it Hera} probe findings, for example, with participation in the European Astrobiology Conference. Papers detailing research results will be submitted to professional journals.

\subsection{Education and Public Outreach}
\label{sec:5.4.8}

The interest of the public in the Saturnian system continues to be significant, with much of the credit for the high interest in Saturn due to the extraordinary success of the {\it Cassini}-{\it Huygens} mission. Images from the Saturnian system are regularly featured as the NASA ``Astronomy Picture of the Day'', and continue to attract the interest of the international media. The interest and excitement of students and the general public can only be amplified by a return to Saturn. The {\it Hera} mission will hold appeal for students at all levels.  Education and Public Outreach activities will be an important part of the {\it Hera} mission planning. An EPO team will be created to develop programs and activities for the general public and students of all ages. Additionally, {\it Hera} results and interpretation of the science will be widely distributed to the public through internet sites, leaflets, public lectures, TV and radio programmes, CD and DVDs, museum and planetarium exhibitions, and in popular science magazines and in newspapers.

 


\section{Summary and conclusion}
\label{sec:7}

{\it In situ} exploration of the giant  planets is necessary to further constrain models of solar system formation and chemical/thermal evolution, and the origin and evolution of atmospheres, to provide a basis for comparative studies of the gas and ice giants, and to provide a unique groundtruth for studying extrasolar planetary systems. In addition, the gas and ice giants provide  a laboratory for which the atmospheric chemistries, dynamics, processes, and interiors of all the planets including Earth can be studied. 

Within the deeper, well-mixed atmospheres of the giant planets the most pristine material from the epoch of solar system formation can be found, providing clues to the local chemical and physical conditions existing at the time and location at which the planet formed. In particular, measurements of noble gas abundances are needed to understand the formation and evolution of the giant planets. In addition to the absolute abundance of noble gases, the ratio of the noble gas abundances will strongly constrain how, when, and where the giant planets formed. Perhaps most important, the abundance of helium will indicate the extent to which helium/hydrogen phase separation has occurred in the deep interior of the giant planets. A Saturn entry probe can determine whether the enhancement of heavy noble gases found in Jupiter by {\it Galileo} is a general feature of all the giant planets, and measurements of the Saturn helium abundance can be contrasted with the Jupiter helium abundance measured by the {\it Galileo} probe, leading to a self-consistent theory for the thermal evolution of both Jupiter and Saturn. This theory will then present a ground truth to calibrate theories of the evolution of giant planet formation, including exoplanets. 

The key goal of a Saturn entry probe mission is to measure the abundances of the noble gases He, Ne, Ar, Kr, Xe and their isotopes, the heavier elements C, N, and S, key isotope ratios $^{14}$N/$^{15}$N, $^{12}$C/$^{13}$C and D/H, and disequilibrium species such as CO, PH$_3$, AsH$_3$, GeH$_4$ which act as tracers of internal processes, and can be achieved by a Saturn entry probe reaching 10 bars. 

Comprising a single probe accompanied by a Carrier Relay Spacecraft, the {\it Hera} Saturn atmospheric entry probe has been proposed as an ESA medium class (M--class) mission. {\it Hera} will measure the noble gas, chemical, and isotopic compositions, processes, and dynamics of Saturn's upper atmosphere, providing science essential for elucidating the origin, formation and thermal and chemical evolution of Saturn, the giant planets, and the solar system, and to provide ground truth measurement for extrasolar planet science. {\it Hera} will probe far below regions accessible to remote sensing, well into the cloudforming region of the troposphere to locations where the most important cosmogenically abundant species well mixed. Along the probe descent, {\it Hera} will provide in situ tracking of Saturn's atmospheric dynamics including zonal winds, waves, convection and turbulence, and measurements of atmospheric pressure and temperature, and the location, density, and composition of the upper cloud layers. 

By leading to an improved understanding of the processes by which giant planets formed, including the composition and properties of the local solar nebula at the time and location of giant planet formation, {\it Hera} will extend the legacy of the {\it Galileo} and {\it Cassini} missions by further addressing the creation, formation, and chemical, dynamical, and thermal evolution of the giant planets, the entire solar system including Earth and the other terrestrial planets, and formation of other planetary systems.

\begin{acknowledgement}
O.M. acknowledges support from CNES. This work has been carried out thanks to the support of the A*MIDEX project (n\textsuperscript{o} ANR-11-IDEX-0001-02) funded by the ``Investissements d'Avenir'' French Government program, managed by the French National Research Agency (ANR).
\end{acknowledgement}

\clearpage

\begin{table}[h]
\caption{Science Priorities. Priority 1 measurements focus only on questions related to Saturn's origin. Science Priorities 2 and 3 address questions related to the structure of Saturn's atmosphere and its formation conditions.}
\label{table1}
\centering
\scriptsize
\resizebox{1\textwidth}{!}{\begin{tabular}{|p{13.5cm}|}
\hline
\bf Priority 1 measurements\\
\hline
\begin{itemize}
\item 1.1 The atmospheric fraction of He/H$_2$ with 2\% accuracy on the measurement (same accuracy as Galileo). A firm measurement of the He abundance is needed to constrain Saturn's interior;
\item 1.2 The abundances on the chemically inert noble gases Ne, Xe, Kr and Ar with 10\% accuracy on the measurement (uncertainties close to those in solar abundances). These elements constitute excellent tracers for the materials in the subreservoirs existing in the PSN;
\item 1.3 The vertical profiles of elemental enrichments in cosmogenically abundant species C, N and S. C/H, N/H and S/H should be sampled with accuracies better than $\pm$ 5\% (same accuracy as {\it Galileo}). The precise measurement of these species provides clues regarding the disk's thermodynamic conditions at the epoch of Saturn's formation.
\end{itemize}
\\
\hline
\bf Priority 2 measurements\\
\hline
\begin{itemize}
\item 2.1 The isotopic ratios in hydrogen (D/H), helium $^3$He/$^4$He, carbon ($^{12}$C/$^{13}$C) and nitrogen ($^{14}$N/$^{15}$N), to determine the key reservoirs for these species (e.g., delivery as N$_2$ or NH$_3$ vastly alters the $^{14}$N/$^{15}$N ratio in the giant planet's envelope). $^3$He/$^4$He should be sampled with an accuracy of $\pm$ 3\% (same as for the Galileo measurement). D/H, $^{12}$C/$^{13}$C, $^{14}$N/$^{15}$N should be analyzed in the main host molecules with an accuracy of the order of $\pm$5\%;
\item 2.2 Continuous measurements of atmospheric temperature and pressure throughout the descent to study (i) stability regimes as a function of depth though transition zones (e.g., radiative-convective boundary); (ii) accelerations; and (iii) the influence of wave perturbations, vertical winds and cloud formation on the vertical temperature profile. Continuous measurement of the conductivity profile to aid in understanding Saturnian lightning.
\end{itemize}
\\
\hline
\bf Priority 3 measurements\\
\hline
\begin{itemize}
\item 3.1 The isotopic ratios in Ne, Ar, Kr and Xe should be measured with accuracy better than $\pm$1\%, to give further constraints on the formation conditions of Saturn in the PSN. $^{16}$O/$^{18}$O and $^{17}$O/$^{18}$O with accuracy better than $\pm$ 1\%, should be sampled in order to investigate possible O isotopic variations throughout the solar system;
\item 3.2 The vertical distributions of minor species to study vertical motions (e.g., NH$_3$, H$_2$S, H$_2$O, PH$_3$, AsH$_3$, GeH$_4$ etc) should be measured from the tropopause to below the condensate clouds. P/H, As/H and Ge/H should be sampled with accuracy better than $\pm$10\% (uncertainties close to solar abundances);
\item 3.3 Measurements of the vertical structure and properties of Saturn's cloud and haze layers; including determinations of the particle optical properties, size distributions, number and mass densities, opacity, shapes and, potentially, their composition;
\item 3.4 Determination of the vertical variation of horizontal winds during the descent.  This includes a study of the depth of the zonal wind fields, and first measurements of middle atmospheric winds;
\item 3.5 Thermal profile and heat budget in the atmosphere;
\item 3.6 Measure accurate photometric light curve to probe for oscillations of the planet;
\item 3.7 Global atmospheric dynamics at equatorial to mid-latitudes from cloud tracking.
\end{itemize}
\\
\hline
\end{tabular}}
\end{table}

\clearpage

\begin{table}[h]
\caption{Suite of scientific instruments.
\label{table2}}
\centering
\begin{tabular}{|p{5cm}|p{1.5cm}|p{5cm}|}
\hline
\bf Instrument 							& \bf Location 		& \bf Measurement											\\
\hline
Mass Spectrometer (MS)					& Probe			& Elemental and chemical composition							\\
 									& 				& Isotopic composition										\\
 									& 				& High molecular mass organics								\\
Atmospheric Structure Instrument (ASI)		& Probe			& Pressure, temperature, density, molecular weight profile, lightning 		\\
Radio Science Experiment (RSE) 			& Probe and Carrier	& Measure winds, speed and direction							\\
									& 				& Chemical composition 										\\
Nephelometer							& Probe			& Cloud structure, solid/liquid particles							\\
Net-flux radiometer (NFR)					& Probe			& Thermal/solar energy										\\
Camera								& Carrier			& Measure winds and cloud structure							\\
									& 				& Detect oscillation signatures									\\
\hline
\end{tabular}
\end{table}

\clearpage

\begin{table}[h]
\caption{Science Traceability Matrix.
\label{table3}
}
\centering
\scriptsize
\begin{tabular}{|p{1.5cm}|p{1.5cm}|p{1cm}|p{3cm}|p{3cm}|p{1.5cm}|}
\hline
\bf Science Goals		& \bf Science Objectives		& \bf Science Priority			& \bf Science Questions	 	& \bf Scientific Measurements		& \bf Instrument\\
\hline
\multirow{4}{1.5cm}{Understand the formation of the Giant Planets and their roles in the evolution of the solar system} 	& \multirow{4}{1.5cm}{Determine the composition of Saturn's well-mixed atmosphere beneath the clouds} 
& 1.1 & What is the abundance of helium relative to H$_2$? & He/H$_2$ ratio to an accuracy of 2\% & MS \\
 & & 1.2 & What are the well-mixed abundances of the noble gases? & Ne/H, Ar/H, Kr/H, Xe/H to a precision of $\pm$ 10\%  & MS\\
 & & 1.3 & What are the abundance profiles of key cosmogenic species? & C/H, N/H and S/H: $\pm$ 5\% & MS, ASI, RSE/AAbs \\
 & & 2.1, 3.1 & What are the most important reservoirs for main isotopes of H, He helium, nitrogen, carbon, oxygen, neon and heavy noble gases? & $^{14}$N/$^{15}$N, $^{12}$C/$^{13}$C D/H: $\pm$ 5\%, $^3$He/$^4$He: $\pm$ 3\% Ne, Ar, Kr and Xe isotopes: $\pm$ 1\%, $^{18}$O/$^{16}$O, $^{17}$O/$^{16}$O: $\pm$ 1\% & MS, ASI\\
\hline
\multirow{6}{1.5cm}{Understand Giant Planet atmospheric circulation, the processes by which energy is transferred outwards from their interior, and the structure of the cloud layers}  & \multirow{6}{1.5cm}{Determine the compositional, thermal, and dynamical structure of Saturn's atmosphere} & 2.2 & What is the vertical structure of Saturn's atmospheric temperatures and stability? & Pressure: $\pm$ 1\%, temperature: $\pm$ 1 K from the upper atmosphere to 10 bar & ASI \\
 & & 3.4 & How do atmospheric winds and wave phenomena vary as a function of depth? & Profile of descent probe telemetry Doppler frequencies Zonal Winds:  $\pm$1 m/s from 0.1--10 bar  & RSE/DWE
Camera \\
 & & 3.2 & How do convective motions and vertical mixing shape the vertical distribution of chemical species? & Vertical profiles of NH$_3$, H$_2$S, H$_2$O, PH$_3$, AsH$_3$, GeH$_4$, CO: $\pm$10\%  & MS, ASI\\
 & & 3.3 & What is the vertical structure, composition and properties of Saturn's cloud and haze layers? & Particle optical properties, size distributions, number and mass densities, opacity, shapes, and composition  & Nephelo-meter, Camera\\
 & & 3.5 & What is the radiative energy balance of the atmosphere? & Up \& down visible flux: $\sim$0.4--5$\mu$m; up \& down IR flux: 4--50$\mu$m; $\lambda$/$\Delta\lambda$$\sim$0.1--100, $\Delta$Flux $\sim$0.5 Wm$^{-2}$  & NFR\\
 & & 3.6 & Does Saturn have acoustic oscillation modes? & Long-term global photometry of the planet  & Camera\\
\hline
\end{tabular}
\end{table}

\clearpage

\begin{table}[h]
\caption{Summary of Parameters for {\it Hera} Science Instruments.\label{Inst}}
\centering
\begin{tabular}{|p{4.5cm}|p{1cm}|p{2.5cm}|p{1.5cm}|p{1.5cm}|p{1.5cm}|}
\hline
								& Mass	& Size								& Power Requirement				& Data Rate		& Data Volume 			\\
\hline
Carrier Camera						& 7.5 kg	& 30$\times$14$\times$12 cm$^3$			& $\sim$12W						& 100 Mb/s		& 6.8 Mbytes per image	\\
Probe Mass Spectrometer				& 16 kg	& 24.5$\times$14.5$\times$22.9 cm$^3$		& $\sim$68W						& $\sim$2 kb/s		& 10.7 Mbit			\\
Probe Atmospheric Structure Inst.		& 1.5 kg	& 20$\times$20$\times$20 cm$^3$			& $\sim$10W						& $\sim$250 b/s	& 1.35 Mbit			\\
Radio Science						& 1.5 kg	& $\sim$4 cm diam $\times$ 14cm length		& $\sim$3W (warmup)				& $\sim$2 b/s		& 10.8 kbit			\\
Net Flux Radiometer					& 2.4 kg	& 11.3$\times$14.4$\times$27.9 cm$^3$		& $\sim$6.3W (peak) $\sim$4.7W (basic)	& $\sim$55 b/s		& 297 kbit				\\
Nephelometer						& 2.3 kg	& 									& $\sim$3W						& $\sim$150 b/s	& 810 kbit				\\
\hline
\end{tabular}
\end{table}

\clearpage





\begin{table}[h]
\caption{List of filters.\label{filters}}
\centering
\begin{tabular}{|p{1.5cm}|p{2cm}|p{5cm}|}
\hline
Name		& Wavelength		& Science													\\
\hline
MT2			& 727 $\pm$ 10 nm	& Dynamics at the haze layers; cloud and hazes vertical distribution		\\
CB2			& 760 $\pm$ 10 nm	& Dynamics at the lower clouds; cloud and hazes vertical distribution		\\
MT3			& 889 $\pm$ 10 nm	& Dynamics at the haze layers; cloud and hazes vertical distribution		\\
CB3			& 940 $\pm$ 20 nm	& Dynamics at the lower clouds; cloud and hazes vertical distribution		\\
UV/VIO		& 380--420 nm		& Cloud and hazes vertical distribution							\\
BLUE		& 400--500 nm		& Outreach; cloud and hazes vertical distribution; color				\\
RED			& 650--750 nm		& Outreach; cloud and hazes vertical distribution; color				\\
GREEN		& 500--600 nm		& Outreach; cloud and hazes vertical distribution; color				\\
\hline
\end{tabular}
\end{table}

\clearpage

\begin{table}[h]
\caption{Current TRL of the camera's elements.\label{TRL_camera}}
\centering
\begin{tabular}{|p{3cm}|p{1cm}|p{3.5cm}|p{3cm}|}
\hline
Sub-system L1				& TRL	& Heritage								& Remarks								\\
\hline	
Camera					& 3		& $Corot$, $Rosetta$/NAC					& 										\\	
Filter wheel assembly		& 3		& Soho/coronograph, Lasco, $Herschel$/Spire/FTS	& Long experience on mechanisms				\\	
Focal plane assembly		& 3		& $Corot$, $Euclid$, $Odin$					&										\\
Assembly detector			& 5		& $JUICE$									& Fully qualified by time needed for this program	\\
Electronic box				& 3		& $Herschel$/Spire/FTS, Corot					& 										\\	
\hline
\end{tabular}
\end{table}

\clearpage

\begin{table}[h]
\caption{Entry System Mass Estimates.\label{Mass_estimate}}
\centering
\scriptsize
\begin{tabular}{|p{3.2cm}|p{1cm}|p{1cm}|}
\hline
Entry Flight Path Angle (EFPA), $\degree$	& -8		& -19			\\
\hline	
									& \multicolumn{2}{c|}{Mass, kg}	\\
Entry System (total mass)					& 216	& 200			\\
Deceleration module						& 92.5	& 76.5			\\
Forebody TPS (HEEET)					& 40		& 24				\\
Afterbody TPS							& 10.5	& 10.5			\\
Structure								& 18.3	& 18.3			\\
Parachute								& 8.2		& 8.2				\\
{Separation Hardware}					& 6.9 	& 6.9				\\
Harness								& 4.3		& 4.3				\\
Thermal Control						& 4.4		& 4.4				\\
Descent Module						& {123.5}	& {123.5}	\\
Communication							& 13		& 13				\\
C\&DH Subsystem						& 18.4	& 18.4			\\
Power Subsystem						& 19.8	& 19.8			\\
Structure								& 30		& 30				\\
Harness								& 9.1		& 9.1				\\
Thermal Control						& 4.3		& 4.3				\\
Science Instrument						& 28		& 28				\\
{Separation Hardware}					& 0.9		& 0.9				\\
\hline
\end{tabular}
\end{table}	
Note. Deceleration of module 1m diameter aeroshell, 36 km/s inertial velocity, 10$\degree$ latitude. {The descent module mass estimate, except for the Science Instruments, is $\sim$5\% percent higher than that of {\it Galileo} Probe due to the presence of a heavier power subsystem. Substantial} mass savings are likely when the descent system structure is adjusted for reduction in scale as well as entry $g$-load.  {\it Galileo} design-to $g$-load was 350.  Saturn probe entry g-load with 3-sigma excursions will be less than 150 $g$'s.

\clearpage

\begin{table}[h]
\caption{Entry $g$-loading, TPS mass comparison between HEEET and HCP, and recession mass loss for the limiting entry conditions.\label{g-load}}
\centering
\begin{tabular}{|p{1cm}|p{1cm}|p{1cm}|p{1cm}|p{1cm}|p{1cm}|p{1cm}|p{1cm}|p{1cm}|p{1cm}|p{1.2cm}|}
\hline
 & Inertial velocity (Km/s)	& Geoc. latitude	& Entry FPA ($\degree$)	& Entry Mass (kg)	& Ballistic Coeff. (kg/m$^2$)	& Entry $g$-load ($g$'s) & HEEET Mass (kg) & Carbon Phenolic Mass (kg) & Mass loss from Recession(kg) & TPS Mass loss/Entry Mass \\
\hline	
1		& 36.0		& 10.0		& -8.0		& 220		& 269		& 29			& 39.3		& 60.8		& 2.7			& 1.2\%		\\
2		& 36.0		& 10.0		& -19.0		& 220		& 269		& 131		& 23.8		& 33.9		& 2.6			& 1.2\%		\\
3		& 36.0		& 0.0			& -8.0		& 200		& 245		& 29			& 29.1		& 44.3		& 1.7			& 0.8\%		\\
4		& 36.0		& 0.0			& -19.0		& 200		& 245		& 127		& 18.7		& 27.1		& 1.6			& 0.8\%		\\
\hline
\end{tabular}
\end{table}

 \clearpage

\begin{table}[h]
\caption{$\Delta$V and propellant budget.\label{DeltaV}}
\centering
\begin{tabular}{|p{2.5cm}|p{1cm}|p{1cm}|p{1cm}|p{1.5cm}|p{2cm}|p{1.5cm}|}
\hline
Manoeuvres					& $\Delta$V	& Margin on $\Delta$V	& Isp (s)	& M final (kg)	& M initial (kg)	& Propellant mass (kg)	\\
\hline	
Launcher dispersion correction		& 10			& 5\%				& 294	& 1271.9		& 1276.6		& 4.6					\\
VGAM and EGAM nav allocation	& 30			& 0\%				& 294	& 1258.8		& 1271.9		& 13.2				\\
DSM before EGA				& 190		& 5\%				& 294	& 1174.6		& 1258.8		& 84.1				\\
Retargeting at Saturn before release	& 15			& 100\%				& 294	& 1162.5		& 1174.6		& 12.2				\\
Avoidance after probe release		& 80			& 5\%				& 294	& 896.0		& 922.6		& 26.5				\\
							&			&					& 		& \multicolumn{2}{c|}{GNC (9kg) +100\% margin (kg)}	& 18			\\
							&			&					&		& \multicolumn{2}{c|}{Residuals mass (kg)}			& 3			\\
							&			&					&		& \multicolumn{2}{c|}{Total propellant mass (kg)}		& 161.7		\\
\hline
\end{tabular}
\end{table}

\clearpage

\begin{table}[h]
\caption{Mass estimates.\label{mass2}}
\centering
\begin{tabular}{|p{4.5cm}|p{1cm}|}
\hline
{\it Hera} mass budget				&  (kg)	\\
\hline	
Probe							& 200	\\
Payload on carrier					& 8		\\
Bus electronics						& 48		\\
Communications					& 55		\\
GNC sensors/actuators				& 25		\\
Mechanisms for probe separation		& 11		\\
Solar Array						& 213	\\
Carrier batteries (primary+secondary)	& 85		\\
Bi-propellant Propulsion S/S			& 44		\\
Thermal Control					& 31		\\
Harness							& 55		\\
Structure							& 154	\\
Nominal Dry Mass at Launch			& 929	\\
System Margin 20\%					& 186	\\
\bf Total Dry Mass at Launch			& \bf 1115	\\
MON/MMH Propellant				& 162	\\
\bf Total Wet Mass at Launch			& \bf 1277	\\
\bf Total Launched (with adapter)		& \bf 1387	\\
\hline
\end{tabular}
\end{table}

\clearpage

\begin{table}[h]
\caption{Work Breakdown Structure for Hera Science Instruments.
\label{table_consortia}}
\centering
\scriptsize
\begin{tabular}{|p{0.5cm}p{0.5cm}|p{4.5cm}|p{2.5cm}|p{4cm}|}
\hline
 \multicolumn{3}{|l}{\bf Instrument} 					& \bf Lead 						& \bf Support										\\
\hline
1.0  	&  & \bf Cameras (on Carrier)						& O. Mousis, PI (FR)					& L. Fletcher, Co-PI (UK); R. Hueso (ES); F.-X. Schmider (FR) 	\\
\hline
	& 1.1 		& Camera optics \& mechanics 		& P. Levacher (FR) 					& 												\\
	& 1.2 		& CMOS chip \& Electronics 		& A. Holland (UK) 					& J. Endicott (UK); M. Leese (UK)						\\
	& 1.3 		& Filter Wheels 				& R. Hueso (ES) 					& C. Ortega (ES); M. A. Carrera (ES)						\\
	& 1.4 		& Electronics box 				& P. Levacher (FR)					&  												\\
\hline
2.0  	&  & \bf Probe Mass Spectrometer (MS)				& P. Wurz, PI (CH)					& J. H. Waite, Co-PI (USA); A. Morse (UK)			 	\\
\hline
	& 2.1 		& TOF-MS, MS Swiss element		& P. Wurz (CH)						& 												\\
	& 2.2 		& GSES, MS US element			& J. H. Waite (USA)					&  												\\
	& 2.3 		& RGS, MS UK element			& A. Morse (UK)					& S. Sheridan (UK)									\\
\hline
3.0  	&  & \bf Probe Atmospheric Structure Investigation (ASI)	& F. Ferri, PI (IT)				& A. Colaprete, Co-PI (USA); G. Fischer (AUT)	 			\\
\hline
	& 3.1 		& Accelerometers (ACC)					& 						& 												\\
	& 3.2 		& Pressure sensors (PPI)					& 						&  												\\
	& 3.3 		& Temperature Sensors (TEM)				& 						& 												\\
	& 3.4 		& Atmospheric Electricity Package (AEP)		& 						&  												\\
	& 3.5 		& ASI Processor (DPU)					& 						& 												\\
\hline
4.0  	&  & \bf Radio Science (Probe and Carrier)					& D. Atkinson, PI (USA)		& T. Spilker (USA)									\\
\hline
	& 4.1 		& Doppler Wind Experiment				& D. Atkinson (USA)			& M. Bird (DE)										\\
	& 4.2 		& Atmospheric UHF, Absorption/NH$_3$ abundance	 & D. Atkinson (USA)		& T. Spilker (USA) 									\\
\hline		
5.0  	&  & \bf Probe Net Flux Radiometer (NFR)						& M. Amato, PI (USA)		& S. Aslam (USA); C. Nixon (USA)					\\
\hline		
	& 5.1 		& Instrument: optics, electronics, mechanical		& S. Aslam (USA)					& M. Amato (USA)						\\
	& 5.2 		& Detector  (Germany) and rad hard ROIC (USA)	& E. Kessler (DE)					& M. Amato (USA)						\\
	& 5.3 		& Filters									& S. Calcutt (UK)					& 									\\
\hline		
6.0  	&  & \bf Probe Nephelometer									& Daphne Stam, PI (NL)		& J.-B. Renard, (FR); O. Munoz (ES);  D. Banfield (USA)	\\
\hline		
	& 6.1 		& Light Optical Aerosol Counter (LOAC)		& J.-B. Renard (FR)					&  										\\
	& 6.2 		& PAVO Optics	 						& C. Keller (NL)						& F. Snik (NL) 								\\
	& 6.3 		& PAVO Detector \& Elect.					& D. Stam (NL)						& 										\\
\hline		
\end{tabular}
\end{table}

%





\clearpage

\clearpage

\clearpage

\clearpage

\clearpage

\clearpage

\clearpage

\clearpage

\clearpage

\clearpage

\clearpage

\clearpage

\begin{thebibliography}{}

\bibitem{Asmar10} Asmar, S.W., D.H. Atkinson, G. Weaver, and N. Yu.: Science-Quality Oscillators for Deep Space Probes, 7th International Planetary Probe Workshop, Barcelona, Spain (2010).

\bibitem{Asmar04} Asmar, S.~W., Atkinson, D.~H., Bird, M.~K., Wood, G.~E.\ 2004.\ Ultra-stable oscillators for planetary entry probes.\ Planetary Probe Atmospheric Entry and Descent Trajectory Analysis and Science {\bf 544}, 131--134 (2004). 

\bibitem{Atkinson98} Atkinson, D.~H., Pollack, J.~B., Seiff, A.:\ The Galileo probe Doppler wind experiment: Measurement of the deep zonal winds on Jupiter.\ Journal of Geophysical Research {\bf 103}, 22911--22928 (1998).

\bibitem{Atkinson97} Atkinson, D.~H., Ingersoll, A.~P., Seiff, A.:\ Deep winds on Jupiter as measured by the Galileo probe.\ Nature {\bf 388}, 649--650 (1997). 

\bibitem{Atkinson96} Atkinson, D.~H., Pollack, J.~B., Seiff, A.:\ Galileo Doppler Measurements of the Deep Zonal Winds at Jupiter.\ Science {\bf 272}, 842--843 (1996). 

\bibitem{Atkinson89} Atkinson, D.~H.:\ Measurement of Planetary Wind Fields by Doppler Monitoring of an Atmospheric Entry Vehicle.\ Ph.D.~Thesis (1989). 

\bibitem{Atreya99} Atreya, S.~K., Wong, M.~H., Owen, T.~C., Mahaffy, P.~R., Niemann, H.~B., de Pater, I., Drossart, P., Encrenaz, T.: A comparison of the atmospheres of Jupiter and Saturn: deep atmospheric composition, cloud structure, vertical mixing, and origin.\ Planetary and Space Science {\bf 47}, 1243--1262 (1999).

\bibitem{Balsiger07} Balsiger, H., and 49 colleagues.:\ ROSINA - Rosetta Orbiter Spectrometer for Ion and Neutral Analysis.\ Space Science Reviews {\bf 128}, 745--801 (2007). 

\bibitem{Boese80} Boese, R.~W., Twarowski, R.~J., Gilland, J., Hassig, R.~E., Brown, F.~G.:\ The infrared radiometer on the sounder probe of the Pioneer Venus mission.\ IEEE Transactions on Geoscience and Remote Sensing {\bf 18}, 97--100 (1980). 

\bibitem{Bird05} Bird, M.~K., and 14 colleagues.:\ The vertical profile of winds on Titan.\ Nature {\bf 438}, 800--802 (2005). 

\bibitem{Bird97} Bird, M.~K., Heyl, M., Allison, M., Asmar, S.~W., Atkinson, D.~H., Edenhofer, P., Wohlmuth, R., Iess, L., Tyler, G.~L., Plettemeier, D.:\ The Huygens Doppler Wind Experiment.\ Huygens: Science, Payload and Mission {\bf} 1177, 139 (1997).

\bibitem{CW01} Chambers, J.~E., Wetherill, G.~W.:\ Planets in the asteroid belt.\ Meteoritics and Planetary Science {\bf 36}, 381--399 (2001). 

\bibitem{Durry02} Durry, G., Hauchecorne, A., Ovarlez, J., Ovarlez, H., Pouchet, I., Zeninari, V.: In situ Measurement of H$_2$O and CH$_4$ with Telecommunication Laser Diodes in the Lower Stratosphere: Dehydration and Indication of a Tropical Air Intrusion at Mid-Latitudes.\ J. Atm. Chem. {\bf 43(3)}, 175--194 (2002). 

\bibitem{Dyudina13} Dyudina, U.~A., Ingersoll, A.~P., Ewald, S.~P., Porco, C.~C., Fischer, G., Yair, Y.\ Saturn's visible lightning, its radio emissions, and the structure of the 2009--2011 lightning storms.\ Icarus {\bf 226}, 1020--1037 (2013). 

\bibitem{Dyudina07} Dyudina, U.~A., Ingersoll, A.~P., Ewald, S.~P., Porco, C.~C., Fischer, G., Kurth, W., Desch, M., Del Genio, A., Barbara, J., Ferrier, J.:\ Lightning storms on Saturn observed by Cassini ISS and RPWS during 2004 2006.\ Icarus {\bf 190}, 545--555 (2007). 

\bibitem{F11} Fletcher, L.~N., Baines, K.~H., Momary, T.~W., Showman, A.~P., Irwin, P.~G.~J., Orton, G.~S., Roos-Serote, M., Merlet, C.:\ Saturn's tropospheric composition and clouds from Cassini/VIMS 4.6-5.1 {$\mu$}m nightside spectroscopy.\ Icarus {\bf 214}, 510--533 (2011). 

\bibitem{F09} Fletcher, L.~N., Orton, G.~S., Teanby, N.~A., Irwin, P.~G.~J..: Phosphine on Jupiter and Saturn from Cassini/CIRS.\ Icarus {\bf 202}, 543--564 (2009). 

\bibitem{Folkner06} Folkner, W.~M., and 16 colleagues.:\ Winds on Titan from ground-based tracking of the Huygens probe.\ Journal of Geophysical Research (Planets) {\bf 111}, E07S02 (2006). 

\bibitem{Folkner98} Folkner, W.~M., Woo, R., Nandi, S.:\ Ammonia abundance in Jupiter's atmosphere derived from the attenuation of the Galileo probe's radio signal.\ Journal of Geophysical Research {\bf 103}, 22847--22856 (1998). 

\bibitem{Folkner97} Folkner, W.~M., Preston, R.~A., Border, J.~S., Navarro, J., Wilson, W.~E., Oestreich, M.:\ Earth-based radio tracking of the Galileo probe for Jupiter wind estimation.\ Science {\bf 275}, 644--646 (1997). 

\bibitem{Fulchignoni05} Fulchignoni, M., and 42 colleagues.:\ In situ measurements of the physical characteristics of Titan's environment.\ Nature {\bf 438}, 785--791 (2005). 

\bibitem{Fulchignoni02} Fulchignoni, M., and 27 colleagues.:\ The Characterisation of Titan's Atmospheric Physical Properties by the Huygens Atmospheric Structure Instrument (Hasi).\ Space Science Reviews {\bf 104}, 397--434 (2002). 

\bibitem{GM11} Garc{\'{\i}}a-Melendo, E., P{\'e}rez-Hoyos, S., S{\'a}nchez-Lavega, A., Hueso, R.: Saturn's zonal wind profile in 2004-2009 from Cassini ISS images and its long-term variability.\ Icarus {\bf 215}, 62--74 (2011).

\bibitem{G05} Gomes, R., Levison, H.~F., Tsiganis, K., Morbidelli, A.: Origin of the cataclysmic Late Heavy Bombardment period of the terrestrial planets.\ Nature {\bf 435}, 466--469 (2005). 


\bibitem{Hansen74} Hansen, J.~E., Travis, L.~D.:\ Light scattering in planetary atmospheres.\ Space Science Reviews {\bf 16}, 527--610 (1974). 

\bibitem{Keller07} Keller, H.~U., and 68 colleagues.:\ OSIRIS -- The Scientific Camera System Onboard Rosetta. Space Science Reviews {\bf 128}, 433--506 (2007). 

\bibitem{Mahaffy12} Mahaffy, P.~R., and 84 colleagues.:\ The Sample Analysis at Mars Investigation and Instrument Suite.\ Space Science Reviews {\bf 170}, 401--478 (2012).


\bibitem{Mousis14} Mousis, O., and 50 colleagues.:\ Scientific rationale for Saturn's in situ exploration.\ Planetary and Space Science {\bf 104}, 29--47 (2014). 

\bibitem{Niemann02} Niemann, H.~B., and 18 colleagues.:\ The Gas Chromatograph Mass Spectrometer for the Huygens Probe.\ Space Science Reviews {\bf 104}, 551--590 (2002). 

\bibitem{Niemann96} Niemann, H.~B., and 12 colleagues.:\ The Galileo Probe Mass Spectrometer: Composition of Jupiter's Atmosphere.\ Science {\bf 272}, 846--849 (1996). 

\bibitem{Orton98} Orton, G.~S., and 16 colleagues: Characteristics of the Galileo probe entry site from Earth-based remote sensing observations. Journal of Geophysical Research 
{\bf 103}, 22791--22814 (1998). 

\bibitem{Owen99} Owen, T., Mahaffy, P., Niemann, H.~B., Atreya, S., Donahue, T., Bar-Nun, A., de Pater, I.: A low-temperature origin for the planetesimals that formed Jupiter.\ Nature {\bf 402}, 269--270 (1999). 

\bibitem{Quiligan14} Quilligan, G., et al.: A 0.18$\mu$m CMOS Thermopile Readout ASIC Immune to 50 Mrad Total Ionizing Dose (Si) and Single Event Latchup to 174 MeV-cm$^2$/mg. International Workshop on Instrumentation for Planetary Missions (IPM-2014), Greenbelt, MD (2014).

\bibitem{Ragent92} Ragent, B., Privette, C.~A., Avrin, P., Waring, J.~G., Carlston, C.~E., Knight, T.~C.~D., Martin, J.~P.:\ Galileo Probe Nephelometer Experiment.\ Space Science Reviews {\bf 60}, 179--201 (1992).

\bibitem{R09} Read, P.~L., Conrath, B.~J., Fletcher, L.~N., Gierasch, P.~J., Simon-Miller, A.~A., Zuchowski, L.~C.: Mapping potential vorticity dynamics on saturn: Zonal mean circulation from Cassini and Voyager data.\ Planetary and Space Science {\bf 57}, 1682--1698 (2009). 

\bibitem{Renard15b} Renard, J.-B., and 32 colleagues:\ LOAC: a small aerosol optical counter/sizer for ground-based and balloon measurements of the size distribution and nature of atmospheric particles -- Part 2: First results from balloon and unmanned aerial vehicle flights.\ Atmospheric Measurement Techniques Discussions {\bf 8}, 1261-1299 (2015b). 

\bibitem{Renard15a} Renard, J.-B., and 32 colleagues:\ LOAC: a small aerosol optical counter/sizer for ground-based and balloon measurements of the size distribution and nature of atmospheric particles -- Part 1: Principle of measurements and instrument evaluation.\ Atmospheric Measurement Techniques Discussions {\bf 8}, 1203--1259 (2015a). 

\bibitem{Renard10} Renard, J.-B., Berthet, G., Salazar, V., Catoire, V., Tagger, M., Gaubicher, B., Robert, C.\ 2010.\ In situ detection of aerosol layers in the middle stratosphere.\  Geophysical Research Letters {\bf 37}, L20803 (2010). 

\bibitem{Rietjens15} Rietjens, J. H. H., van Harten, G., Bekkers, D., et al.: Performance of spectrally modulated polarimetry II: Data reduction and absolute polarization calibration of a prototype SPEX satellite instrument, Appl. Optics, submitted.

\bibitem{SL91} Sanchez-Lavega, A., Colas, F., Lecacheux, J., Laques, P., Parker, D., Miyazaki, I.: The Great White SPOT and disturbances in Saturn's equatorial atmosphere during 1990.\ Nature {\bf 353}, 397--401 (1991). 

\bibitem{Seiff98} Seiff, A., Kirk, D.~B., Knight, T.~C.~D., Young, R.~E., Mihalov, J.~D., Young, L.~A., Milos, F.~S., Schubert, G., Blanchard, R.~C., Atkinson, D.:\ Thermal structure of Jupiter's atmosphere near the edge of a 5-{$\mu$}m hot spot in the north equatorial belt.\ Journal of Geophysical Research {\bf 103}, 22857--22890 (1998).

\bibitem{Seiff92} Seiff, A., Knight, T.~C.~D.:\ The Galileo Probe Atmosphere Structure Instrument.\ Space Science Reviews {\bf 60}, 203--232 (1992). 

\bibitem{Seiff80} Seiff, A., Kirk, D.~B., Young, R.~E., Blanchard, R.~C., Findlay, J.~T., Kelly, G.~M., Sommer, S.~C.:\ Measurements of thermal structure and thermal contrasts in the atmosphere of Venus and related dynamical observations - Results from the four Pioneer Venus probes.\ Journal of Geophysical Research {\bf 85}, 7903-7933 (1980). 

\bibitem{Snik09} Snik, F., Karalidi, T., Keller, C.~U.:\ Spectral modulation for full linear polarimetry.\ Applied Optics {\bf 48}, 1337--1346 (2009). 

\bibitem{Soman14} Soman, M., Holland, A.~D., Stefanov, K.~D., Gow, J.~P., Leese, M., Pratlong, J., Turner, P.:\ Design and characterisation of the new CIS115 sensor for JANUS, the high resolution camera on JUICE.\ Society of Photo-Optical Instrumentation Engineers (SPIE) Conference Series {\bf 9154}, 915407 (2014). 

\bibitem{Sromovsky98} Sromovsky, L.~A., Collard, A.~D., Fry, P.~M., Orton, G.~S., Lemmon, M.~T., Tomasko, M.~G., Freedman, R.~S.:\ Galileo probe measurements of thermal and solar radiation fluxes in the Jovian atmosphere.\ Journal of Geophysical Research {\bf 103}, 22929--22977 (1998).

\bibitem{vZ98} von Zahn, U., Hunten, D.~M., Lehmacher, G.: Helium in Jupiter's atmosphere: Results from the Galileo probe helium interferometer experiment.\ Journal of Geophysical Research {\bf 103}, 22815--22830 (1998). 

\bibitem{Webster11} Webster, C.~R., Mahaffy, P.~R.:\ Determining the local abundance of Martian methane and its $^{13}$C/$^{12}$C and D/H isotopic ratios for comparison with related gas and soil analysis on the 2011 Mars Science Laboratory (MSL) mission.\ Planetary and Space Science {\bf 59}, 271--283 (2011).

\bibitem{Wong04} Wong, M.~H., Mahaffy, P.~R., Atreya, S.~K., Niemann, H.~B., Owen, T.~C.: Updated Galileo probe mass spectrometer measurements of carbon, oxygen, nitrogen, and 
sulfur on Jupiter.\ Icarus {\bf 171}, 153--170 (2004).

\bibitem{W12} Wurz, P., Abplanalp, D., Tulej, M., Lammer, H.:\ A neutral gas mass spectrometer for the investigation of lunar volatiles.\ Planetary and Space Science {\bf 74}, 264--269 (2012). 

\bibitem{Zarnecki04} Zarnecki, J.~C., Ferri, F., Hathi, B., Leese, M.~R., Ball, A.~J., Colombatti, G., Fulchignoni, M.:\ In-flight performance of the HASI servo accelerometer and implications for results at Titan.\ ESA-SP-544, 71--76 (2004).







\end{thebibliography}
\end{document}